\documentclass[review]{elsarticle}

\usepackage[table]{xcolor}  
\usepackage{multirow}
\usepackage{mathtools}
\usepackage{amsmath}
\usepackage{amsmath,amssymb,amsthm}
\usepackage{graphicx}
\usepackage{subfig}
\usepackage[framemethod=tikz]{mdframed}
\usepackage{soul}
\usepackage{siunitx}

\usepackage{commath}
\usepackage{url}

\usepackage{caption}
\usepackage{esvect}
\usepackage{booktabs} 
\usepackage{gensymb}
\usepackage{algcompatible}
\usepackage{algorithm}
\usepackage[noend]{algpseudocode}
\usepackage{setspace}

\makeatletter
\renewcommand{\ALG@beginalgorithmic}{\small}
\makeatother

\usepackage{tablefootnote}

\makeatletter
\def\therule{\makebox[\algorithmicindent][l]{\hspace*{.5em}\vrule height .75\baselineskip depth .25\baselineskip}}%

\newtoks\therules
\therules={}
\def\appendto#1#2{\expandafter#1\expandafter{\the#1#2}}
\def\gobblefirst#1{
  #1\expandafter\expandafter\expandafter{\expandafter\@gobble\the#1}}%
\def\LState{\State\unskip\the\therules}
\def\LStatex{\Statex\unskip\the\therules}
\def\pushindent{\appendto\therules\therule}%
\def\popindent{\gobblefirst\therules}%
\def\printindent{\unskip\the\therules}%
\def\printandpush{\printindent\pushindent}%
\def\popandprint{\popindent\printindent}%

\algdef{SE}[WHILE]{While}{EndWhile}[1]
  {\printandpush\algorithmicwhile\ #1\ \algorithmicdo}
  {\popandprint\algorithmicend\ \algorithmicwhile}%
\algdef{SE}[FOR]{For}{EndFor}[1]
  {\printandpush\algorithmicfor\ #1\ \algorithmicdo}
  {\popandprint\algorithmicend\ \algorithmicfor}%
\algdef{S}[FOR]{ForAll}[1]
  {\printindent\algorithmicforall\ #1\ \algorithmicdo}%
\algdef{SE}[LOOP]{Loop}{EndLoop}
  {\printandpush\algorithmicloop}
  {\popandprint\algorithmicend\ \algorithmicloop}%
\algdef{SE}[REPEAT]{Repeat}{Until}
  {\printandpush\algorithmicrepeat}[1]
  {\popandprint\algorithmicuntil\ #1}%
\algdef{SE}[IF]{If}{EndIf}[1]
  {\printandpush\algorithmicif\ #1\ \algorithmicthen}
  {\popandprint\algorithmicend\ \algorithmicif}%
\algdef{C}[IF]{IF}{ElsIf}[1]
  {\popandprint\pushindent\algorithmicelse\ \algorithmicif\ #1\ \algorithmicthen}%
\algdef{Ce}[ELSE]{IF}{Else}{EndIf}
  {\popandprint\pushindent\algorithmicelse}%
\algdef{SE}[PROCEDURE]{Procedure}{EndProcedure}[2]
   {\printandpush\algorithmicprocedure\ \textproc{#1}\ifthenelse{\equal{#2}{}}{}{(#2)}}%
   {\popandprint\algorithmicend\ \algorithmicprocedure}%
\algdef{SE}[FUNCTION]{Function}{EndFunction}[2]
   {\printandpush\algorithmicfunction\ \textproc{#1}\ifthenelse{\equal{#2}{}}{}{(#2)}}%
   {\popandprint\algorithmicend\ \algorithmicfunction}%
\makeatother

\makeatletter
\def\BState{\State\hskip-\ALG@thistlm}
\makeatother

\usepackage{setspace}

\usepackage{breakcites}
\usepackage{enumerate}
\usepackage{makecell}

\usepackage{float}
\usepackage{listings}
\usepackage{breqn}
\usepackage{tikz}
\usetikzlibrary{trees}
\usepackage{paralist}
\usepackage{ragged2e}
\usepackage[framemethod=tikz]{mdframed}

\usetikzlibrary{positioning,arrows.meta}

\definecolor{arrowblue}{RGB}{98,145,224}

\usepackage{smartdiagram}
\usepackage[margin=3cm]{geometry}
\newtheorem{defn}{Definition}[section]

\usepackage{footnote}

\makeatletter
\def\BState{\State\hskip-\ALG@thistlm}
\makeatother

\usepackage{setspace}

\usepackage{breakcites}
\usepackage{enumerate}
\usepackage{makecell}

\usepackage{float}
\usepackage{listings}

\usepackage{tikz}
\usetikzlibrary{trees}
\usepackage{paralist}
\usepackage{ragged2e}
\usepackage[framemethod=tikz]{mdframed}

\usetikzlibrary{positioning,arrows.meta}

\definecolor{arrowblue}{RGB}{98,145,224}

\usepackage{soul}

\usepackage[colorinlistoftodos]{todonotes}

\usepackage{lineno,hyperref}
\usepackage{cleveref}
\usepackage{setspace}
\journal{Journal of Information Sciences}

\begin{document}

\begin{frontmatter}

\title{Efficient Privacy Preservation of Big Data for Accurate Data Mining}



\author[mymainaddress,mysecondaryaddress]{M.A.P.~Chamikara
	\corref{mycorrespondingauthor}}
\cortext[mycorrespondingauthor]{Corresponding author}
\ead{pathumchamikara.mahawagaarachchige@rmit.edu.au}

\author[mymainaddress]{P.~Bertok}
\author[mysecondaryaddress]{D.~Liu}
\author[mysecondaryaddress]{S.~Camtepe}
\author[mymainaddress]{I.~Khalil}

\address[mymainaddress]{RMIT University, Australia}
\address[mysecondaryaddress]{CSIRO Data61, Australia}

\begin{abstract}
\begin{mdframed}[backgroundcolor=green!50,rightline=false,leftline=false]
\centering 
The published article can be found at \url{https://doi.org/10.1016/j.ins.2019.05.053}
\end{mdframed}
Computing technologies pervade physical spaces and human lives, and produce vast amount of data that is available for analysis. However, there is a growing concern that potentially sensitive data may become public if the collected data are not appropriately sanitized before being released for investigation. Although there are more than a few privacy-preserving methods available, they are not efficient, scalable or have problems with data utility, and/or privacy. This paper addresses these issues by proposing an efficient and scalable nonreversible perturbation algorithm, PABIDOT, for privacy preservation of big data via optimal geometric transformations. PABIDOT was tested for efficiency, scalability, resistance, and accuracy using nine datasets and five classification algorithms. Experiments show that PABIDOT excels in execution speed, scalability, attack resistance and accuracy in large-scale privacy-preserving data classification when compared with two other, related privacy-preserving algorithms. 
\end{abstract}


\end{frontmatter}


\section{Introduction}
\label{introduction}

Recent advances in computer technologies have drastically increased the amount of data collected from cyber, physical and human worlds. Data collection at large scale can make sense only if they are actionable and they can be used in decision making ~\cite{witten2016data}. Data mining helps at this point by investigating unsuspected relationships in the data and providing useful insights to the data owners. Moreover, such capabilities may often need to be shared with external parties for further analysis.  In this process, various kinds of information may be revealed, which can lead to a privacy breach. The ability to share information while preventing the disclosure of personally identifiable information (PII) becomes an important aspect of information privacy, and it is one of the most significant technical, legal, ethical and social challenges. In fact, various governmental and commercial organizations collect vast amounts of user data, among others individual credit information, health, financial status, and personal preferences. Social networking, banking and healthcare systems are examples of systems that handle such private information~\cite{chamikaraprocal}, and they often overlook privacy due to indirect use of private information. There are other information systems that use massive amounts of sensitive private information (also called big data) for modeling and prediction of human-related phenomena such as crimes ~\cite{helbing2015saving}, epidemics ~\cite{jalili2017information} and grand challenges in social physics ~\cite{capraro2018grand}. Hence, privacy preservation (a.k.a. sanitization) ~\cite{vatsalan2017privacy} 
can become a very complex problem and requires robust solutions~\cite{wen2018scheduling}.

Privacy-preserving data mining (PPDM) offers the possibility of using data mining methods without disclosing private information. PPDM approaches include data perturbation (data modification) ~\cite{chen2005random, chen2011geometric} and encryption ~\cite{kerschbaum2017searchable}. Cryptographic methods are renowned for securing data.  Literature provides many examples where PPDM effectively utilize cryptographic methods ~\cite{li2017privacy}. For example, we can find applications of homomorphic encryption in the domains including but not limited to e-health, cloud computing and sensor networks~\cite{zhou2015ppdm}. Secure sum, secure set union, scalar product and set intersection are few other operations that can be used as building blocks in distributed data mining ~\cite{clifton2002tools}.  However, due to their high computational complexity, they cannot provide sufficient data utility~\cite{gai2016privacy} and are impractical for PPDM. Data perturbation is known to have lower computational complexity compared to cryptographic methods for privacy preservation ~\cite{chamikaraprocal}. Data perturbation maintains individual record confidentiality by applying a systematic modification to the data elements in a database ~\cite{chamikaraprocal}. The perturbed dataset is often indistinguishable from the original dataset, e.g. age data maps to a reasonable number so that a third party cannot differentiate between original and perturbed age. Examples of perturbation techniques include adding noise to the original data (additive perturbation) ~\cite{muralidhar1999general}, applying rotation to the original data using a random rotation matrix (random rotation)  ~\cite{chen2005random}, applying both rotation and translation to the original data using a random rotation matrix and a random translation matrix (geometric perturbation) ~\cite{chen2011geometric}, and randomizing the outputs of user responses using some random algorithm (randomized response) ~\cite{dwork2014algorithmic}. A major disadvantage of these techniques is that they can not process high volumes of data efficiently, e.g. random rotation and geometric perturbation consume a considerable amount of time to provide good results while enforcing sufficient privacy~\cite{chen2005random,chen2011geometric}. Additive perturbation takes less time but provides a lower privacy guarantee ~\cite{okkalioglu2015survey}. The existing methods have issues in maintaining a proper balance between privacy and utility. 

It is essential to define the effectiveness of a privacy-preserving approach using a privacy model, and identify the limitations of private information protection and disclosure  ~\cite{chamikaraprocal}. $K-anonymity$, $l-diversity$, $(\alpha, k)-anonymity$, $t-closeness$ are some of the previous privacy models, and they show vulnerability to different attacks, e.g. minimality, composition and foreground knowledge attacks~\cite{chamikaraprocal}. Differential privacy (DP) is a privacy model known to render maximum privacy by minimizing the chance of individual record identification~\cite{dwork2014algorithmic}. Local differential privacy (LDP), achieved by input perturbation ~\cite{dwork2014algorithmic}, allows full or partial data release to analysts   ~\cite{kairouz2014extremal} by randomizing the individual instances of a database~\cite{tang2017privacy}. Global differential privacy (GDP), also called the trusted curator model, allows analysts only to request the curator to run queries on the database. The curator applies carefully calibrated noise to the query results to provide differential privacy  ~\cite{dwork2014algorithmic,kairouz2014extremal}.  However, GDP and LDP fail for small datasets, as accurate estimation of the statistics shows poor results when the number of tuples is small.  Although differential privacy has been studied thoroughly, only a few viable solutions exist towards full/partial data release using LDP. Most of these are solutions for categorical data such as RAPPOR~\cite{erlingsson2014rappor} and Local, Private, Efficient Protocols for Succinct Histograms~\cite{qin2016heavy}.  DP's solid, theoretically appealing foundation for privacy protection has limited the practicality of implementing efficient solutions towards big data. Furthermore, existing LDP algorithms include a significant amount of noise addition (i.e. randomization), resulting in low data utility. Accordingly, utility and privacy often appear as conflicting factors, and improved privacy usually entails reduced utility.

The main contribution of this paper is a new Privacy preservation Algorithm for Big Data Using Optimal geometric Transformations (PABIDOT). PABIDOT is an irreversible input perturbation mechanism with a new privacy model ($\Phi-separation $) which facilitates full data release.  We prove that $\Phi-separation $ provides an empirical privacy guarantee against the data reconstruction attacks. PABIDOT is substantially faster than comparable methods; it sequentially applies random axis reflection, noise translation, and multidimensional concatenated subplane rotation followed by randomized expansion and random tuple shuffling for further randomization. Randomized expansion is a novel method to increase the positiveness or the negativeness of a particular data instance. PABIDOT's memory overhead is comparatively close to other solutions, and it provides better attack resistance, classification accuracy, and excellent efficiency towards big data. We tested PABIDOT by using nine generic datasets retrieved from the UCI machine learning data repository\footnote{http://archive.ics.uci.edu/ml/index.php} and the OpenML machine learning data repository\footnote{https://www.openml.org},  the results were compared against two alternatives: random rotation perturbation (RP) ~\cite{chen2005random} and geometric perturbation (GP) ~\cite{chen2011geometric}, which are known to provide high utility in terms of privacy-preserving classification. Our study shows that PABIDOT always ends up with approximately optimal perturbation. PABIDOT produces the best empirical privacy possible by determining the globally optimal perturbation parameters adhering to $\Phi-separation$ for the dataset of interest. The source code of the PABIDOT project is available at $https://github.com/chamikara1986/PABIDOT$.

The rest of the paper is organized as follows. Section \ref{literature} provides a summary of existing related work. The technical details of PABIDOT are described in Section \ref{proposedalgo}. Section \ref{proposedalgo} further presents the basic flow of PABIDOT which we refer as PABIDOT\_basic for convenience. The efficiency optimization of PABIDOT is discussed in Section \ref{pabeffi}. At the end of Section \ref{pabeffi}, the main algorithm (PABIDOT) with optimized efficiency is introduced. Section \ref{pabeffi} presents experimental settings and provides a comparative analysis of performance and resistance of PABIDOT.  The results are discussed in Section \ref{discussion}, and the paper is concluded in Section \ref{conclusion}.

\section{Literature Review}
\label{literature}
Privacy protection of individuals has become a challenging task with the proliferation of Internet-enabled consumer technologies. Literature shows different approaches to find solutions towards this challenge. While some approaches concentrate on increasing the awareness~\cite{buccafurri2016threat}, others try to employ different techniques to enforce individual privacy~\cite{wei2018autoprivacy}. Above all, the massive volumes of big data introduce many challenges to privacy preservation~\cite{cuzzocrea2015privacy}. Although the security and privacy concerns of big data are not entirely new,  they require attention due to the specifics of the environments and dynamics put forward by the devices used~\cite{kieseberg2018security}. The advancements of these environments and the diversity of devices always introduce increased complexity and make security and privacy preservation complex.    To counter the diversified challenges and complexities, three different technological approaches can be observed: disclosure control, privacy-preserving data mining (PPDM) and privacy-enhancing technologies ~\cite{torra2017data}. Attribute-based encryption, controlling access via authentication, temporal and location-based access control and employing constraint-based protocols are some mechanisms that are used for improving the privacy of systems in dynamic environments~\cite{chamikaraprocal}.  Among the various approaches to privacy-preserving data mining, data perturbation is often preferred due to its simplicity and efficiency ~\cite{aldeen2015comprehensive}. Both input and output perturbation are often used:  output perturbation is based on noise addition and rule hiding while input perturbation is conducted either by noise addition~\cite{muralidhar1999general} or multiplication ~\cite{chamikaraprocal}. Input perturbation can be divided further into unidimensional perturbation and multidimensional perturbation~\cite{okkalioglu2015survey}. Additive perturbation ~\cite{muralidhar1999general}, randomized response~\cite{dwork2014algorithmic}, swapping ~\cite{hasan2016effective} and microaggregation~\cite{torra2017fuzzy} are examples of unidimensional input perturbation, whereas condensation~\cite{aggarwal2004condensation}, random rotation ~\cite{chen2005random}, geometric perturbation~\cite{chen2011geometric}, random projection ~\cite{liu2006random}, and hybrid perturbation are multidimensional~\cite{aldeen2015comprehensive}.

In additive perturbation, random noise is added to the original data in such a way that the underlying statistical properties of the attributes are preserved. A significant problem with this approach is low utility of the resulting data ~\cite{agrawal2000privacy}. Additionally, effective noise reconstruction techniques developed in response can significantly reduce the level of privacy ~\cite{okkalioglu2015survey}. Randomization techniques such as randomized response is another approach ~\cite{dwork2014algorithmic}, e.g. randomizing the responses of interviewees in order to preserve the privacy of respondents. Due to the high randomization of input data, randomization techniques such as randomized response often provide high privacy whereas the utility in terms of estimating statistics or conducting analyses can be low ~\cite{dwork2014algorithmic}. Microaggregation is based on confidentiality rules that allow the publication of micro datasets. It divides the dataset into clusters of $k$ elements and replaces the values in each cluster with the centroid of the cluster. Microaggregation to a single variable (univariate microaggregation) is vulnerable to transparency attacks when the published data includes information about the protection method and its parameters ~\cite{torra2017fuzzy}. Multivariate microaggregation has also been proposed, but it is complex and has been proven to be NP-hard  ~\cite{torra2017fuzzy}. In condensation, the input dataset is divided into multiple groups of a pre-defined size in such a way that the difference between records in a particular group is minimal,  and a certain level of statistical information about the different records is maintained in each group. Then sanitized data are generated using uniform random distribution based on the eigenvectors which are generated using the eigendecomposition of the characteristic covariance matrices of each homogeneous group ~\cite{aggarwal2004condensation}. Condensation has a significant shortcoming in that it may degrade the quality of data significantly.

Random rotation perturbation, geometric data perturbation, and random projection perturbation are three types of matrix multiplicative ~\cite{okkalioglu2015survey} methods. In random rotation, the original data matrix is multiplied using a random rotation matrix that has the properties of an orthogonal matrix. The application of rotation is repeated until the algorithm converges at the desired level of privacy ~\cite{chen2005random}.  In geometric data perturbation, a random translation matrix is incorporated in the process of perturbation in order to enhance privacy. The method accompanies three components: rotation perturbation, translation perturbation, and distance perturbation ~\cite{chen2011geometric}. The main idea of random projection perturbation is to project data from high-dimensional space to a randomly chosen low-dimensional subspace ~\cite{liu2006random}. Due to the isometric nature of transformations, random rotation perturbation, geometric data perturbation, and random projection perturbation are capable of preserving the distances between tuples in a dataset ~\cite{chen2005random,chen2011geometric,liu2006random}. Accordingly, they provide high utility w.r.t. classification and clustering.  In hybrid perturbation, both matrix multiplicative and matrix additive properties are used, which is quite similar to geometric perturbation ~\cite{aldeen2015comprehensive}. These algorithms have high computational complexity and are time-consuming, which make them unsuitable to work with big datasets.

Due to its explicit notion of strong privacy guarantee, differential privacy has attracted much attention. Although LDP permits full or partial data release, and the analysis of privacy-protected data~\cite{dwork2014algorithmic,kairouz2014extremal}, LDP algorithms are still at a fundamental stage when it comes to the privacy preservation of real-valued numerical data. The complexity of selecting the domain of randomization with respect to a single data instance is still a challenge~\cite{erlingsson2014rappor}. In  GDP, the requirement of a trusted curator who enforces differential privacy by applying noise or randomization can be considered as a primary issue~\cite{dwork2014algorithmic}.  The fundamental mechanisms used to obtain differential privacy include Laplace mechanism, Gaussian mechanism ~\cite{dwork2014algorithmic}, geometric mechanism, randomized response, and staircase mechanisms ~\cite{kairouz2014extremal}. The necessity of a trusted third party in GDP and the application of extremely high noise in LDP are inherent shortcomings that directly affect the balance between privacy and utility of these practical, differentially private approaches.

Many previously proposed privacy preservation methods, including data perturbation, perform poorly with high dimensional datasets. The necessary computing resources grow fast as the number of attributes and number of instances increase even though the performance is good for low dimensional data. This quality is often referred to as ``The Dimensionality Curse" ~\cite{chamikaraprocal}. Large datasets also provide extra information to attackers, as higher dimensions help in utilizing background knowledge to identify individuals ~\cite{bettini2015privacy}. 

\label{literaturelast}

Most of the privacy-preserving algorithms have problems with balancing privacy and utility. Data privacy focuses on the difficulty of estimating the original data from the sanitized data, while utility concentrates on preserving application-specific properties/information ~\cite{aggarwal2015privacy}. A generic way of measuring the utility of a privacy-preserving method is to investigate perturbation biases ~\cite{wilson2008protecting}. Data perturbation bias means that the result of a query on the perturbed data is significantly different from the result generated for the same query on the original data. Wilson et al. have examined different data perturbation methods against various bias measures ~\cite{wilson2008protecting}, namely, A, B, C, D, and Data Mining (DM) bias. Type A bias occurs when the perturbation of a given attribute causes summary measures to change. Type B bias is the result of the perturbation changing the relationships between confidential attributes, while in case of Type C bias the relationship between confidential and non-confidential attributes changes. Type D bias means that the underlying distribution of the data was affected by the sanitization process. If Type DM bias exists, data mining tools will perform less accurately on the perturbed data than they would on the original dataset. It has been noted that privacy preservation mechanisms decrease utility in general, and finding a trade-off between privacy protection and data utility for big data is an important issue ~\cite{xu2015privacy}.

In the literature, there is a dearth of efficient privacy preservation methods that provide reliable data utility and are scalable enough to handle the rapidly growing data. Existing methods also have problems with levels of uncertainty, biases and low level of resistance.  To address the issues presented by big data, there is an urgent need for methods that are scalable, efficient and robust. New methods should overcome the aforementioned weaknesses of the existing PPDM methods and provide solutions towards large-scale privacy preserving data mining.

\section{Proposed Algorithm: PABIDOT}
\label{proposedalgo}
\begin{figure}[H]
\centering
\scalebox{0.8}{
\includegraphics[width=1\textwidth, trim=0cm 0cm 0cm 0cm]{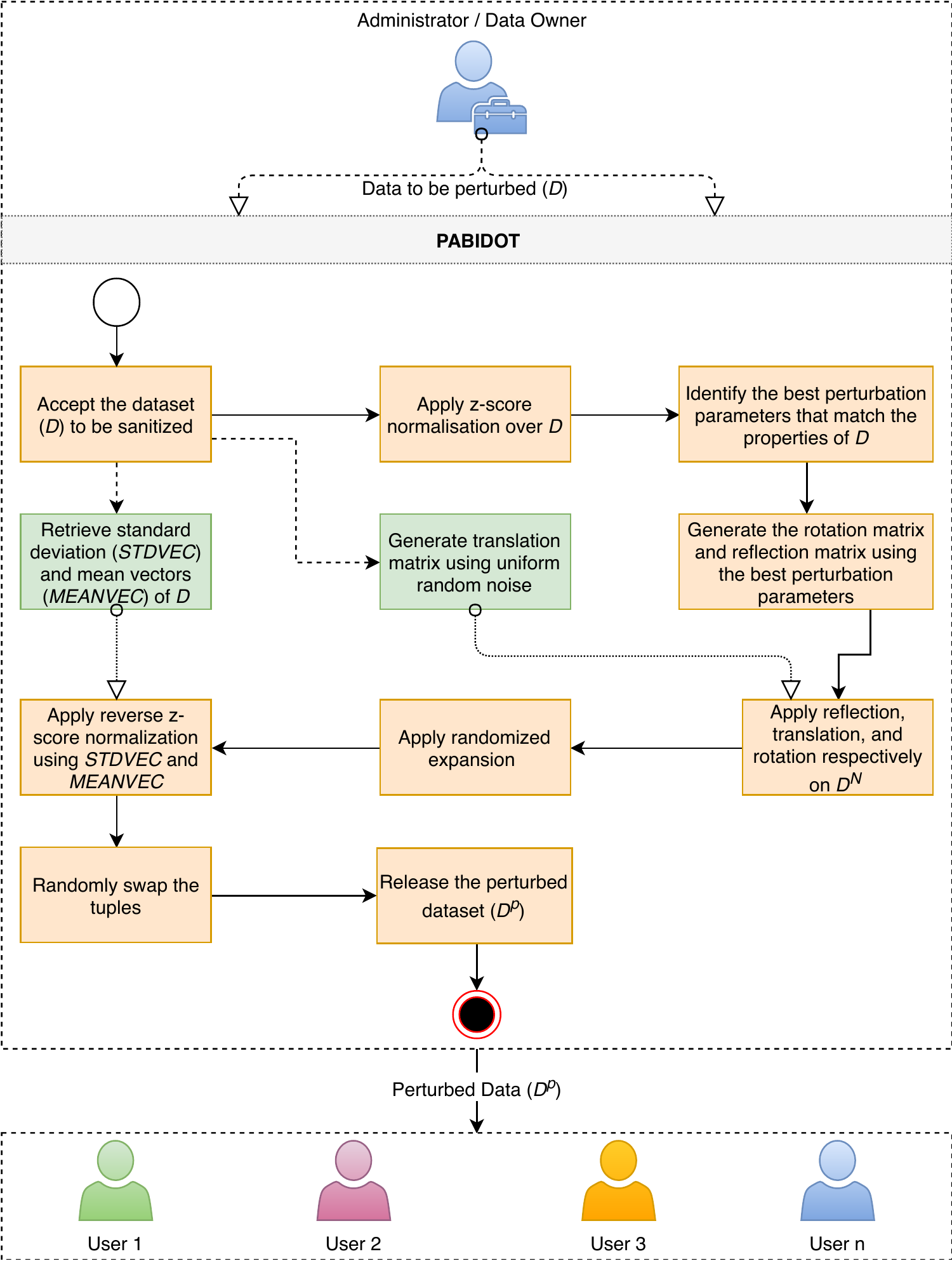}
	}
\caption{Basic flow and the architecture of PABIDOT. In this setting, the data owner is considered to be the trusted curator who owns the original dataset. The owner is located in the local edge of a cloud computing scenario. The orange boxes represent the main steps of the algorithm whereas the green boxes represent the intermediate data generative steps which support the appropriate main steps.}


\label{algoflow}
\end{figure}

PABIDOT perturbs a dataset by using multidimensional geometric transformations, reflection, translation, and rotation followed by randomized expansion (a new noise addition mechanism which is explained in Section \ref{randexp}) and random tuple shuffling. Figure \ref{algoflow} shows the basic flow and architecture of the proposed perturbation algorithm. Based on the proposed privacy model called $\Phi-separation$, the algorithm aims at optimum privacy protection in terms of protection against data reconstruction attacks. PABIDOT achieves this by selecting the best possible perturbation parameters based on the properties of the input dataset.  Figure \ref{algoflow} also demonstrates the position of PABIDOT in a privacy-preserving big data release scenario.  PABIDOT assumes that the original data can be accessed only by the owner/ the administrator of that dataset. There can be complementary releases of the perturbed versions of the original dataset. The original dataset will not be released to a third party users under any circumstances.

\paragraph*{\textbf{Rationale and technical novelty}} PABIDOT applies geometric transformations with optimal perturbation parameters and increases randomness using randomized expansion followed by a random tuple shuffle. It defines privacy in such a way that the resulting dataset has an optimal difference compared to the original dataset as a result of the privacy model ($\Phi-separation$) used by PABIDOT. This property helps in minimizing the search space and finding the best possible perturbation for a particular dataset. Consequently, efficiency and reliability of PABIDOT in big data perturbation increase while providing better resistance to data reconstruction. Figure \ref{algoflow} and Algorithm \ref{parallelalgo} depict the proposed perturbation algorithm, and Table \ref{notations} provides a summary of the notations used in Algorithm \ref{exhuativeapproach} and Algorithm \ref{parallelalgo}. As shown, the original dataset and the standard deviation to the normal random noise provided under randomized expansion are the only inputs to Algorithm \ref{parallelalgo}, and the perturbed dataset is its only output.

\paragraph*{\textbf{Data matrix (D)}} The dataset to be perturbed is represented as a matrix ($D$) of size $m \times n$ where the columns represent the attributes ($n$ attributes), and rows represent the records ($m$ records). For example, personal information of a patient can be represented as a  record which may have attributes such as age, weight, height, and gender. The data matrix is assumed to contain numerical data only.
 
\begin{equation}
D=
\begin{bmatrix}
a_{11} & \dots & a_{1k} & \dots & a_{1n}\\
a_{21} & \dots & a_{2k} & \dots & a_{2n}\\
\vdots & \ddots & \vdots & \ddots & \vdots \\ 
a_{m1} & \dots & a_{mk} & \dots & a_{mn}\\
\end{bmatrix}_{m\times n}
\end{equation}
In the process of perturbation, the data matrix is subjected to multidimensional geometric composite transformations. During these transformations, a record (row) in the data matrix will be considered as a point in the multi-dimensional Cartesian coordinate system.

\paragraph*{\textbf{Multidimensional isometric transformations}} Geometric translation, rotation and reflection are considered to be
isometric transformations in the $\ \textit{n-dimensional} $ space. A transformation $\ T: R^n\rightarrow R^n $ is said to be
isometric, if it preserves the distances so that ~\cite{maruskin2012essential}
\begin{equation}
|T(A)-T(B)|=|A-B|,\quad\forall~A,B\in R^n.
\end{equation}

All matrices and Cartesian points are represented in homogeneous coordinate form in order to consider all the transformations in matrix multiplication
form. A homogeneous coordinate point in $\ \textit{n-dimensional}$ space can be written as an $\ (n+1) $ dimensional position vector $\ (x_1,x_2,\dots,x_n,h)$ with the additional term $\ h \neq 0$. The introduction of homogeneous coordinates enables composite transformation between the coordinates and transformation matrices without having to perform a collection of transformations as a sequential process.  Therefore, multidimensional geometric translation, reflection and rotation can be represented in their generalized forms of $\ (n + 1)\times(n + 1) $ matrices ~\cite{jones2012computer}.

A composite operation is performed when several transformation matrices have to be applied in a particular transformation. If several transformation matrices $\ M_1, M_2, M_3,\dots $ are sequentially applied to a homogeneous matrix $\ X $, the composite operation is given by
\begin{equation}
X'=\dots(M_3(M_2(M_{1}X)))
=\dots M_3\times M_2\times M_1\times X
\end{equation}

\paragraph*{\textbf{Homogeneous data matrix}}  All records in the input data matrix ($D$) are converted to homogeneous coordinates by adding a new column of ones (i.e. $\ h = 1 $) after the $n^{th}$ column. The resulting homogeneous representation of the data matrix is given by Equation \ref{datamathomogeneous}.

The input dataset will first be subjected to z-score normalization ~\cite{kabir2015novel} in order to provide equal weights for all attributes in the transformations. Next, the \textit{n-dimensional} translational matrix is generated according to Equation \ref{translationmat}, in which the translational coefficients are drawn from  random noise with uniform distribution, and n equals to the number of attributes in the input dataset. Due to z-score normalization, the attribute mean becomes 0 while the standard deviation of the dataset becomes 1. Therefore, the noise generated by the uniform random noise function is bounded within $(0,1)$ and follows the inequality $0< t_i(n+1) <1 $ where $t_i(n+1)$ denotes a translational coefficient.

\begin{equation}
D=
\begin{bmatrix}
a_{11} & \dots & a_{1k} & \dots & a_{1n} & 1\\
a_{21} & \dots & a_{2k} & \dots & a_{2n} & 1\\
\vdots & \ddots & \vdots & \ddots & \vdots & \vdots \\ 
a_{m1} & \dots & a_{mk} & \dots & a_{mn} & 1\\
\end{bmatrix}_{m\times (n+1)}
\label{datamathomogeneous}
\end{equation}

\paragraph*{\textbf{n-Dimensional translation matrix}} The $\ \textit{n-dimensional} $ homogeneous translation matrix $\ T_{ND} $ can be derived as shown in Equation \ref{translationmat}  ~\cite{jones2012computer}.
\begin{equation}
	T_{ND}=
	\begin{bmatrix}
	1 & 0 & 0 & \dots & 0  & t_{1(n+1)} \\
	0 & 1 & 0 & \dots & 0 & t_{2(n+1)} \\
	0 & 0 & 1 & \dots & 0  & t_{3(n+1)} \\
	\vdots & \vdots & & \ddots & & \vdots \\
	0 & 0 & 0 & \dots & 1  & t_{(n)(n+1)} \\
	0 & 0 & 0 & \dots & 0 & 1 \\
	\end{bmatrix}_{(n+1)\times (n+1)}
\label{translationmat}
\end{equation}

\paragraph*{\textbf{n-Dimensional reflection matrix}} Next, for each of the $n$ axes  PABIDOT generates the corresponding matrix of reflection according to Equation \ref{refmatn1}. The $\ \textit{n-dimensional} $ homogeneous matrix $\ RF_{ND} $ for reflection across axis one can be derived as shown in Equation \ref{reflectionmat}  ~\cite{jones2012computer}.

\begin{equation}
	RF_{ND}=
	\begin{bmatrix}
	1 & 0 & 0 & \dots & 0 & 0 & 0 \\
	0 & -1 & 0 & \dots & 0 & 0 & 0 \\
	0 & 0 & -1 & \dots & 0 & 0 & 0 \\
	\vdots & \vdots & & \ddots & & \vdots & \vdots \\ 
	0 & 0 & 0 & \dots & -1 & 0 & 0 \\
	0 & 0 & 0 & \dots & 0 & -1 & 0 \\
	0 & 0 & 0 & \dots & 0 & 0 & 1 \\
	\end{bmatrix}_{(n+1)\times (n+1)}
\label{reflectionmat}
\end{equation}

The (n+1) axis reflection matrix can be written as shown in Equation \ref{refmatn1}.

\begin{equation}
RF_{\overline{ND}}=
\begin{bmatrix}
-1 & 0 & 0 & \dots & 0 & 0 & 0 \\
0 & 1 & 0 & \dots & 0 & 0 & 0 \\
0 & 0 & 1 & \dots & 0 & 0 & 0 \\
\vdots & \vdots & & \ddots & & \vdots & \vdots \\ 
0 & 0 & 0 & \dots & 1 & 0 & 0 \\
0 & 0 & 0 & \dots & 0 & 1 & 0 \\
0 & 0 & 0 & \dots & 0 & 0 & 1 \\
\end{bmatrix}_{(n+1)\times (n+1)}
\label{refmatn1}
\end{equation}

\paragraph*{\textbf{n-Dimensional rotation matrix}} After creating the matrix of reflection for one of the $n$ number of axes, PABIDOT generates the \textit{n-dimensional} concatenated subplane rotation matrices ($M$ for the current $\theta_i$) using Algorithm \ref{rotmatgenerate} for each $\theta_i$ where ($(0<\theta_i<\pi),\quad\theta_i \notin \left\{ \pi/6, \pi/4, \pi/3, \pi/2, 2\pi/3, 3\pi/4, 5\pi/6  \right\} $)

In order to derive a single matrix that represents the entire $\ n-dimensional $ orientation, the concatenated sub-plane rotation method can be used. Then, the rotations in the plane of a pair of coordinate axes $\ (\hat{x_i}, \hat{x_j})$ for $1\leq i,j \leq n $ can be written as a block matrix as in Equation \ref{rotationmat}  ~\cite{paeth2014graphics}.

\begin{equation}
	R_{ij}(\theta_{ij})=
	\begin{bmatrix}
	1 & \dots & 0 & 0 & \dots & 0 & 0 & \dots & 0 \\
	\vdots & \ddots & \vdots & \vdots & \ddots & \vdots & \vdots & \ddots & \vdots \\
	0 & \dots & cos \theta_{ij} & 0 & \dots & 0 & -sin \theta_{ij} & \dots & 0\\
	0 & \dots & 0 & 1 & \dots & 0 & 0 & \dots & 0 \\
	\vdots & \ddots & \vdots & \vdots & \ddots & \vdots & \vdots & \ddots & \vdots \\
	0 & \dots & 0 & 0 & \dots & 1 & 0 & \dots & 0 \\
	0 & \dots & sin \theta_{ij}& 0 & \dots & 0 &  cos \theta_{ij}  & \dots & 0\\
	\vdots & \ddots & \vdots & \vdots & \ddots & \vdots & \vdots & \ddots & \vdots \\
	0 & \dots & 0 & 0 & \dots & 0 & 0 & \dots & 1 
	\end{bmatrix}_{(n+1)\times (n+1)}
\label{rotationmat}
\end{equation}
Thus $\ \frac{N(N-1)}{2} $ distinct $\ R_{ij}(\theta_{ij}) $ should be concatenated in the
preferred order to produce the final composite \textit{n-dimensional} matrix which can be obtained using Equation \ref{rotmul}  with $\ \frac{N(N-1)}{2} $ degrees of freedom, parameterized by$\ \theta_{ij} $.

\begin{equation}
M=\prod_{i<j} R_{ij}(\theta_{ij})
\label{rotmul}
\end{equation}

\paragraph*{\textbf{Algorithm for generating the multidimensional concatenated sub-plane rotation matrix}} Based on Equations \ref{rotationmat} and \ref{rotmul},  Algorithm \ref{rotmatgenerate}  can be used to generate the multidimensional concatenated subplane rotational matrix of the desired rotation angle. 

The resulting rotation matrix has the properties of an orthogonal matrix where the columns and rows of the  resulting concatenated subplane rotation matrix ($R(\theta)$) are orthonormal, and hence $R(\theta)$ preserves the relationship,  $R(\theta)\times R(\theta)^T=R(\theta)^T\times R(\theta)=I$ where $R(\theta)^T$ be the transpose matrix of $R(\theta)$ and $I$ be the identity matrix.

\begin{algorithm}[H]
	\caption{Algorithm for generating multidimensional concatenated subplane rotation matrix}\label{rotmatgenerate}
	\begin{spacing}{1.2}
		
		\begin{algorithmic}[1]
			\Statex\textbf{Inputs} :
			
			\begin{tabular}{l c l} 
				$n               $ & $\gets $ & number of attributes of the dataset\\
				$\theta $ & $ \gets $ & rotation angle \\
				
			\end{tabular}
			
			\Statex \textbf{Outputs}:
			
			\begin{tabular}{ l c l } 
				$M $ & $\ \gets $ & multidimensional rotation matrix of $ \theta $  
			\end{tabular}			
			\LState $V $ = $ \{1,2,\hdots ,n\} $
			\LState $C$ = $ \{(i,j)| i,j \in V $ and $ i \neq j \}  $
			\LState	$I_n $ = identity matrix of size $n$ 
			\LState	$N $ = $ \binom{n}{2} $ \Comment{Total number of $sin(\theta)$ and $cos(\theta)$ assignments necessary}
            \LState $I3=I_n $
			\For{\texttt{ $ k = 1 $ $ to $ $ N $ }} 
			\LState $  A=\{(i_k,i_k),(j_k,i_k),(i_k,j_k),(j_k,j_k)\} $ where $ \{i_k,j_k \}$  
            \LStatex is the $ k^{th} set$ $of $ $ C$ \Comment{Coordinates of $sin(\theta)$ and $cos(\theta)$ assignments for an instance of $C$}
			\LState $I2$ $= $ $I_{n}$ \Comment{Initializing $I_n$ with the identity matrix of size $n$}
			\LState $I2(i_{k1},j_{k1})$ $= $ $cos(\theta)$ 
			\LStatex where, $(i_{k1},j_{k1})$ is the set number 1 of $ A $
			\LState $I2(i_{k2},j_{k2})$ $= $ $sin(\theta)$ 
 			\LStatex where, $(i_{k2},j_{k2})$ is the set number 2 of $ A $
			\LState $I2(i_{k3},j_{k3})$ $= $ $-sin(\theta)$ 
 			\LStatex where, $(i_{k3},j_{k3})$ is the set number 3 of $ A $
			\LState $I2(i_{k4},j_{k4})$ $= $ $cos(\theta)$ 
 			\LStatex where, $(i_{k4},j_{k4})$ is the set number 4 of $ A $
			\LState $I3=I3\times I2 $ \Comment{Iterative multiplication of $I2$ and $I3$ to form $M$ according to  Equations \ref{rotationmat} and \ref{rotmul}}
			
			\EndFor
			\State 	$\ M=I3 $ \Comment{follows Equations \ref{rotationmat} and \ref{rotmul}}

			\Statex \textbf{End Algorithm}
		\end{algorithmic}
		
	\end{spacing}
\end{algorithm}

\paragraph*{\textbf{Privacy metric for generating the optimal perturbation parameters}} Since the proposed method is based on multidimensional isometric transformations, it is important to use a multi-column privacy metric for evaluating the privacy of the proposed method. Assuming that all attributes of the dataset are equally important, z-score normalization is applied to the data matrix as the initial step in the perturbation process. The higher the privacy of the perturbed data, the more difficult it is to estimate the original data ~\cite{chen2005random}. To extend this idea, the variance of the difference between the perturbed and non-perturbed datasets ($Var(P)$) is considered, the higher the $Var(P)$, the higher the privacy.   Hence,  $Var(P)$ provides a measure of privacy of the perturbed data, or the level of difficulty to estimate the original data without prior knowledge about the original data ~\cite{chen2005random}, which is often called naive inference/estimation ~\cite{chen2005random}. $Var(P)$   has long been used to measure the level of privacy of perturbed data ~\cite{muralidhar1999general}. In the proposed method the attribute which returns the minimum variance for the difference is considered as the  minimum privacy guarantee. If $\ X^p$ is a perturbed data series of attribute $\ X $, the level of privacy of the perturbation method can be measured using $\ Var(P)$, where $P=(X^p-X)$. Therefore, $\ Var(P)$ can be written as in Equation \ref{varp}.
\begin{equation}
Var(P)=Var(p_1,p_2,\dots,p_n)={\frac{1}{n}}\displaystyle\sum_{i=1}^{n}(p_i-\bar{p})^2
\label{varp}
\end{equation}

We propose a new privacy definition called $\Phi-separation $ as follows. Assume we have a dataset $D$ of $m$ instances and each instance having $n$ attributes. We perturb this dataset in $k$ different ways, producing datasets $D_j^p=[d^p_{j1},d^p_{j2},\dots, d^p_{jn}]_{m\times n}$ for $j=1,2,\dots, k$. The perturbed value of an attribute $d$ is denoted as $d^p$. We then calculate the difference between the original data and the perturbed data $D$ - $D_j^p$, which will contain values of ($d_i$ - $d^p_i$) for $i=1,2,\dots, n$. We calculate the variance of each attribute in $D$ - $D_j^p$, and select the minimum of these variances. 

\begin{equation}
\label{phieq0}
\phi_j=min\{[var(d_i-d^p_{ji})]\} \forall j={1,2,\dots,k}
\end{equation}

From the $k$ perturbations, we choose the one that has the largest value of $\Phi$, i.e. we choose the perturbation that produced the most significant difference between the original dataset and the perturbed dataset.

\begin{defn}[$\Phi-separation$]

Given the dataset $D=[d_1,d_2,\dots, d_n]_{m\times n}$ and a perturbation algorithm, we shall denote a perturbed instance of $D$ as $D_j^p=[d^p_{j1},d^p_{j2},\dots, d^p_{jn}]_{m\times n}$ for $j=1,2,\dots, k$, where $k$ represents the number of possible ways of perturbing $D$. Then, the minimum privacy guarantee $\phi_j$ is defined as

\begin{equation}
\label{phieq1}
\phi_j=min\{[var(d_i-d^p_{ji})]\} \forall j={1,2,\dots,k}
\end{equation}

\noindent and the optimal privacy guarantee $\Phi$ is

\begin{equation}
\label{phieq2}
\Phi=max\{\phi_j\}_{j=1}^k
\end{equation}
A perturbed dataset that has the optimal privacy guarantee provides $\Phi-separation$.
\end{defn}

For a perturbed data matrix $\ D^p $ of $\ D $ of size $\ m\times n $, the minimum privacy guarantee ($\phi$) is the minimum of the variances calculated for attributes based on the differences between the original and perturbed z-score normalized values, as shown in Equation \ref{ftheta}. Here $\ \phi $ is the minimum privacy guarantee and $n$ is the total number of attributes in $D$ or $D^p $. 
\begin{equation}
\resizebox{0.5\textwidth}{!}{$\displaystyle
	\phi=min(Var(d_{ij}-d_{ij}^{p})^m_{i=1})^n_{j=1} \qquad \forall d_{ij} \in D, \quad d_{ij}^{p} \in D^p
	$}
\label{ftheta}
\end{equation}
\paragraph*{\textbf{Identifying the best perturbation parameters for $D^N$}} In each iteration the algorithm maximizes the value of $\phi$ to generate $\Phi$ as given in Equation \ref{fthetamax}. With the axis of reflection varying from 1 to $n$ (number of attributes) and the angle of rotation ($\theta$) varying from 0 to 179 degrees, i.e. $0<\theta<\pi$ and $\quad\theta \notin \left\{ \pi/6, \pi/4, \pi/3, \pi/2, 2\pi/3, 3\pi/4, 5\pi/6  \right\} $, the perturbation will return a $n\times \theta$ number of perturbed data matrices: $\ D_{1,1}^p, D_{1,2}^p,\dots, D_{1,179}^p,D_{2,1}^p, D_{2,2}^p,\dots, D_{2,179}^p,\dots,D_{n,1}^p, \\D_{n,2}^p,\dots, D_{n,179}^p  $ with different levels of perturbations. This will form the matrix of local minimum privacy guarantees, $\phi_{i,j}$ as given in Equation \ref{phivals}.

\begin{equation}
\begin{bmatrix}
\phi_{1,1} & \phi_{1,2} & \dots & \phi_{1,179} \\
\phi_{2,1}  & \phi_{2,2} & \dots & \phi_{2,179} \\
\vdots  & \vdots & \ddots & \vdots  \\ 
\phi_{n,1} & \phi_{n,2} & \dots & \phi_{n,179} \
\end{bmatrix}_{n\times 179}
\label{phivals}
\end{equation}
To get the global minimum guarantee ($\phi_j$)values for each angle, we get the minimum of each column as given in Equation \ref{globephi}.
\begin{equation}
\begin{bmatrix}
\phi_{1} & \phi_{2} & \dots & \phi_{179} \\
\end{bmatrix}
\label{globephi}
\end{equation}

The perturbed data matrix with the optimal privacy can be considered as the data matrix that returns the largest value $\Phi$ for the minimum privacy guarantee $ \phi$. Therefore, the largest global minimum privacy guarantee ($\phi_j$) is selected from Equation \ref{globephi} to obtain  $\Phi$ (Equation \ref{fthetamax}). In other words,  the best perturbation parameters will be selected using the highest privacy guarantee $\Phi $, based on the most vulnerable attributes.
\begin{equation}
\Phi=max([[\phi_{j}]_{j=1}^{179}])
\label{fthetamax}
\end{equation}

\paragraph*{\textbf{Generate the rotation matrix and reflection matrix using the best perturbation parameters for $D^N$}}   Next, PABIDOT records the angle of rotation ($\theta_{optimal}$) and axis of reflection ($Rif_{optimal}$) at $\Phi$. Now, the algorithm uses $Rif_{optimal}$ to generate the matrix of reflection according to Equation \ref{refmatn1} and use $\theta_{optimal}$ to generate the matrix of rotation according to Algorithm \ref{rotmatgenerate}. Composite transformation of reflection, translation, and rotation will then be applied to the z-score normalized matrix using the matrices of optimal reflection, translation, and optimal rotation to generate $D^{p2}$.
 
 \label{randexp}

\begin{figure}[H]
	\centering
	\scalebox{0.48}{
	\includegraphics[width=1\textwidth, trim=0cm 0cm 0cm 0cm]{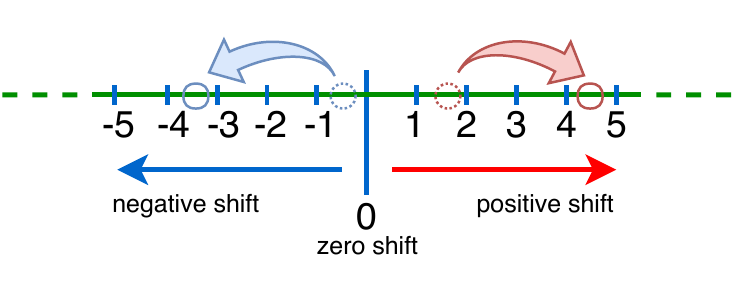}
	}
	\caption{Effect of $Randomized$ $Expansion$. The red arrows of the right-hand side show a positive shift where a calibrated positive random value is added to the positive value to increase the positiveness of the original value. The left-hand side which is represented by the blue arrows show a negative shift where a calibrated negative random value is added to the negative value to increase the negativeness of the original value.}
	\label{shiftnoise}

\end{figure}

  \paragraph*{\textbf{Application of z-score normalization and the transformations: reflection, translation and rotation}} After generating the concatenated subplane rotation matrix, PABIDOT applies the composite transformation of reflection, translation, and rotation on the z-score normalized input data matrix. Rotation is applied after reflection and noise translation. This is because the effect of rotation is proportional to the distance from the origin, and we want to reduce the probability of points close to the origin getting attacked due to weaker perturbation ~\cite{chen2011geometric}. An instance of the application of the transformation to the input data matrix ($D$) in the order of application can be represented using Equation \ref{transformationinstance}.

\begin{equation}
D^{'}=(M\times T_{ND}\times RF_{\overline{ND}}\times (D^N)^T)^T
\label{transformationinstance}
\end{equation}
\paragraph*{\textbf{Randomized Expansion}}  A privacy preserving algorithm is composable if it can keep satisfying the same privacy requirements of the privacy model in use, after repeated independent application of the algorithm~\cite{soria2016big}. To improve  the composability and the randomness  of the perturbation algorithm, noise drawn from random normal distribution (with a mean of 0, and a predefined standard deviation subjected to a default value of 0.3) is added to the data according to a novel approach named as randomized expansion. Here, we introduce noise in such a way that it further enhances the positiveness or the negativeness of a particular value where the zeros are not subjected to any change as depicted in Figure \ref{shiftnoise}.
 
In order to generate the  noise for randomized expansion, first, we generate the sign matrix ($S_{\pm}$) based on the values of $D^{p2}$. A value in $S_{\pm}$ will be 1 if the corresponding element of $D^{p2}$ is greater than 0, it will be 0 if the corresponding element of $D^{p2}$ equals 0, and it will be -1 if the corresponding element of $D^{p2}$ is less than 0. $D^{p2}./abs(D^{p2})$ if $D^{p2}$ is complex. Next, the absolute values of  $D^{p2}$ and absolute values of the random normal noise matrix $(\mathcal{N}(0,\sigma))$ are added together and dot product with  $S_{\pm}$ is calculated as denoted in Equation \ref{plusminus}.

\begin{equation}
\label{plusminus}
\begin{aligned}
D^{p2}=(\left\| D^{p2}\right\|+\left\| \mathcal{N}(0,\sigma)\right\|)\bullet S_{\pm}
\end{aligned}
\end{equation}

 \paragraph*{\textbf{Application of reverse z-score normalization and random tuple swapping}} At this stage, the attributes of the perturbed data matrix will not be within the ranges of the attributes of the original dataset. Therefore, reverse z-score normalization will be applied to the current matrix, using the standard deviations ($STDVEC$) and means of the attributes ($MEANVEC$) of the original dataset. Finally, the tuples of the resulting matrix will be randomly swapped to generate the final perturbed dataset. 
 
 \begin{table}[H]
	\centering
	
	\caption{Summary of the notations used in the proposed algorithm.}  
	\label{notations}
	
	\begin{small}
    \resizebox{0.75\columnwidth}{!}{
		\begin{tabular}{ l l }
			\hline
			{\bfseries Notation} & {\bfseries  Summary/Constraints} \\
			\hline
			$\Phi$ &  $\ \Phi$ is the optimal privacy guarantee \\ 
			
			$\theta_{optimal}$ & $\ \theta_i $ at $\Phi$ \\ 
			
			$Rif_{optimal}$ & optimal axis for $(n-1)$ reflection  selected at $\Phi$\\
			
			$m$& number of tuples\\
			
			$n$& number of attributes\\
			
			$D^{N} $ & z-score normalized dataset of $\ D$\\
            $Cov(D^{N}) $ & covariance matrix of $\ D^N$\\
			
			$TN^{noise}$ & transformation matrix with uniform random noise \\
			
			$RF_{ax} $ &  $\ \textit{n-dimensional} $ reflection matrix for the current axis $ax$ \\
                       & (generated according to Equation \ref{refmatn1}) \\
			
			$\theta_i$ & ($\ 0< \theta_i < \pi $) and $ \theta_i \notin \left\{ \pi/6, \pi/4, \pi/3, \pi/2, 2\pi/3, 3\pi/4, 5\pi/6  \right\}$\\
			
			$M_i $ &  $\ \textit{n-dimensional} $ concatenated subplane rotation  matrix \\
                   & generated using Algorithm \ref{rotmatgenerate} for $\theta_i$ \\
			
			$M\theta $ & $\ \textit{n-dimensional} $ concatenate sub-plane rotation  matrix \\
                       &  generated using Algorithm \ref{rotmatgenerate} for $\theta_{optimal}$ \\
			
			$RF_{optimal} $ & $\ RF_{optimal} $ is the $\ \textit{n-dimensional} $ reflection matrix, \\ 
                            &  generated according to  Equation \ref{refmatn1} for $\ Rif_{optimal}$\\
			
			$STDVEC$ & vector of standard deviations of all the attributes of $\ D$  \\ 
		
			$MEANVEC$ & vector of mean values of all the attributes of $\ D$  \\ 
			\hline
		\end{tabular}
        }
	\end{small} 
\end{table}

\subsection{Overall process of PABIDOT} 
Algorithm \ref{exhuativeapproach} summarizes the overall set of  steps of the basic perturbation algorithm. A summary of notations used in this algorithm is provided in Table \ref{notations}. For convenience, we refer to Algorithm \ref{exhuativeapproach} as PABIDOT\_basic in subsequent sections.

\begin{algorithm}[H]
	\caption{Basic steps of PABIDOT}\label{exhuativeapproach}
	\begin{spacing}{1.2}
		\begin{algorithmic}[1]
			\linespread{1}
			
			\LStatex\textbf{Inputs} :
			
			\begin{tabular}{l c l} 
				$D$  & $\ \gets $ & original dataset\\
                $\sigma$ & $\gets $ & input noise standard deviation (default value=0.3)\\
			\end{tabular}
			
			\LStatex \textbf{Outputs}:
			
			\begin{tabular}{ l c l } 
				$D^{p}$  & $\ \gets $ & perturbed dataset  \\
			\end{tabular}
			\LState $\Phi=0$
			\LState $\theta_{optimal} = 0$  
			\LState $Rif_{optimal}=0$ 
			\LState generate $D^{N} $ \Comment{by applying z-score normalization on $D$}
		
			\LState generate $TN^{noise} $ \Comment{according to Equation \ref{translationmat}, using uniform random noise as the translational coefficients}
			
			\For{\textbf{each} $ax$ in  $\left\{1,2,\dots,n\right\}$ } \Comment{assume there are $n$ number of attributes in $D$}
			\LState generate $RF_{ax} $    \Comment{according to Equation \ref{refmatn1}}
			\For{\textbf{each} $\theta_i $ }
			\LState generate $M_i $ using Algorithm \ref{rotmatgenerate} \Comment{refer Equations \ref{rotationmat}, and  \ref{rotmul}}
			\LState $D^{p1}=(M_i\times TN^{noise}\times RF_{ax}\times (D^N)^T)^T$  \label{step10} \Comment{follows Equation \ref{transformationinstance}}
			\LState $\phi_{i,ax}=min(Var(x_{ij}-p_{ij})_{i=1}^m)_{j=1}^n$  $\forall$ $x_{ij}\in D^N$ \label{step11} and $p_{ij} \in D^{p1}$  \Comment{according to Equation \ref{ftheta}}
			\EndFor			
			\EndFor \label{al2step18}
			\For{\textbf{each} $\theta_i $ }
            \LState   $\ \phi_{i}=min(\phi_{i,{ax}})$  where, $ax$ $\in$  $\left\{1,2,\dots,n\right\}$	 \Comment{follows Equation \ref{globephi}}
			\EndFor			
            \LState $\Phi = max(\phi_{i})$ where, $i$ $\in$  $\left\{1,2,\dots,\theta \right\}$ \Comment{according to Equation \ref{fthetamax}}
            \LState $\theta_{optimal}=\theta_i$  at $\Phi $
			\LState $Rif_{optimal}=ax$  at $\Phi $

			\LState generate $M\theta $  \Comment{according to Algorithm \ref{rotmatgenerate}, using $\theta_{optimal}$}
			
			\LState generate $RF_{optimal} $  \Comment{according to Equation \ref{refmatn1}, using $Rif_{optimal}$}
			\LState$D^{p2}=(M\theta \times TN^{noise}\times RF_{optimal}\times (D^N)^T)^T$ \label{step21} \Comment{follows Equation \ref{transformationinstance}}
			
            \State $D^{p2}=(\left\| D^{p2}\right\|+\left\| \mathcal{N}(0,\sigma)\right\|$)$\bullet S_{\pm}$ \label{stdnc1} \Comment{according to Equation \ref{plusminus}}
			\LState $D^{p} $=$\ D^{p3} \bullet STDVEC$+$MEANVEC$ \Comment{reverse z-score normalization}
            
			\State randomly swap the tuples of $\ D^{p} $ \label{swapstep} \label{step23}

			\Statex \textbf{End Algorithm}
		\end{algorithmic}
	\end{spacing}
\end{algorithm}

\section{Optimizing the efficiency of PABIDOT}
\label{pabeffi}

During the execution of PABIDOT, the composite transformations in steps \ref{step10} and \ref{step11} in Algorithm \ref{exhuativeapproach} need to be applied to the whole dataset for a total of $n\times \theta $ times. This can be infeasible for high dimensional datasets. To accelerate the computations, a method based on the covariance matrix of the input dataset ($D$)  was used to generate $\phi_{i,ax}$, where $\phi_{i,ax}=min(\vv1+diag(RT\times RF\times Cov(D)\times RF\times RT^T)+sum((Cov(D)\times RF \bullet RT)^T)^T)$. This eliminates the necessity of searching through a large number of tuples of a big dataset in each loop to find $\Phi$, as it only needs the $Cov(D)$ of $D$. The steps of this derivation are provided below (Section \ref{distributedproof}).
\begingroup
\setlength\abovedisplayskip{0pt}
\begin{proof}[Poof:]
\label{distributedproof}
\begin{equation}
\begin{aligned}
Var(X-X^p)&=E(X-X^p)^2-[E(X-X^p)]^2\\
&=E(X^2-2XX^p-{X^p}^2)-[(E(X))^2-2E(X)E(X^p)+(E(X^p))^2]\\
&=[E(X^2)-(E(X))^2]+[E({X^p}^2)-(E(X^p))^2]-2[E(XX^p)-E(X)E(X^p)]\\
&=Var(X)+Var(X^p)-2Cov(X,X^p)
\end{aligned}
\end{equation}
Since, X is z-score normalized $\ Var(X)$ is equal to $\ (n-1)/n $. For large number of records $\ (n-1)/n = 1 $. 

Therefore,
\begin{equation}
\begin{aligned}
Var(X-X^p)=1+Var(X^p)-2\times Cov(X,X^p)
\end{aligned}
\end{equation}
Hence,
\begin{equation}
\begin{aligned}
Var(D-D^p)=\vv 1+Var(D^p)-2\times Cov(D,D^p)
\label{vardd}
\end{aligned}
\end{equation}

Let's consider the following table $\ D $ as the table to be perturbed where $\ X, Y, Z $ are the columns of $\ D $,

\begin{center}
	\begin{tabular}{ c c c c  } 
		\toprule
		& X & Y & Z \\
		\midrule
		$\ a_1 =>$ &$\ x_1  $\ & $\ y_1  $\  & $\ z_1  $  \\ 
		$\ a_2 =>$ &$\ x_2  $\ & $\ y_2  $\  & $\ z_2  $  \\ 
		$\ a_3 =>$ &$\ x_3  $\ & $\ y_3  $\  & $\ z_3  $  \\ 
		\bottomrule
	\end{tabular}
\end{center}

The perturbation of a single tuple $a_i$ can be given as $\ a_i^p=RT\times TN\times RF\times a_i^T$ (where, $RT$,$TN$ and $RF$ are rotation, translation and reflection matrices respectively), which is found using

\begin{equation}
\begin{bmatrix}
x^{p}_{1}\\
y^{p}_{1}\\
z^{p}_{1}\\
1
\end{bmatrix}
=
\begin{bmatrix}
a & b & c & 0\\
d & e & f & 0\\
g & h & i & 0\\
0 & 0 & 0 & 1
\end{bmatrix}
\begin{bmatrix}
1 & 0 & 0 & nx\\
0 & 1 & 0 & ny\\
0 & 0 & 1 & nz\\
0 & 0 & 0 & 1
\end{bmatrix}
\begin{bmatrix}
-1 & 0 & 0 & 0\\
0 & 1 & 0 & 0\\
0 & 0 & 1 & 0\\
0 & 0 & 0 & 1
\end{bmatrix}
\begin{bmatrix}
x_{1}\\
y_{1}\\
z_{1}\\
1
\end{bmatrix}
\end{equation}

Assuming that axis 1 has been selected for the $(n-1)$ reflection.

Therefore,

\begin{equation}
\begin{aligned}
x^{p}_1&=-a(x_1+n_x) +b(y_1+n_y)+c(z_1+n_z)=-ax_1 +by_1+cz_1+an_x+bn_y+cn_z\\
x^{p}_2&=-a(x_2+n_x) +b(y_2+n_y)+c(z_2+n_z)=-ax_2 +by_2+cz_2+an_x+bn_y+cn_z\\
x^{p}_3&=-a(x_3+n_x) +b(y_3+n_y)+c(z_3+n_z)=-ax_3 +by_3+cz_3+an_x+bn_y+cn_z\\
\end{aligned}
\label{pertseq}
\end{equation}

and

\[
\begin{bmatrix}
x^{p}_{1}\\
x^{p}_{2}\\
\vdots\\
x^{p}_{n}
\end{bmatrix}
=
-a
\begin{bmatrix}
x_{1}\\
x_{2}\\
\vdots\\
x_{n}
\end{bmatrix}+
b
\begin{bmatrix}
y_{1}\\
y_{2}\\
\vdots\\
y_{n}
\end{bmatrix}+
c
\begin{bmatrix}
z_{1}\\
z_{2}\\
\vdots\\
z_{n}
\end{bmatrix}
+aN_x+bN_y+cN_z
\]

or
\begin{equation}
\begin{aligned}
X^{p}&=-aX+bY+cZ+ \delta E
\end{aligned}
\label{perteqs}
\end{equation}

Where, $\delta E = aN_x+bN_y+cN_z $

So,
\begin{equation}
\begin{aligned}
Cov(X,X^{p})&=Cov(X,(-aX+bY+cZ+ \delta E))\\
&=Cov(X,-aX)+Cov(X,bY)+Cov(X,cZ)\\
&=-aCov(X,X)+bCov(X,Y)+cCov(X,Y)
\end{aligned}
\end{equation}

Similarly,
\begin{equation}
\begin{aligned}
Cov(Y,Y^{p})&=-dCov(Y,X)+eCov(Y,Y)+fCov(X,Z)
\end{aligned}
\end{equation}
\begin{equation}
\begin{aligned}
Cov(Z,Z^{p})&=-gCov(Z,X)+hCov(Z,Y)+iCov(Z,Z)
\end{aligned}
\end{equation}

and

\begin{equation}
Cov(D,D^{p})
=
\begin{bmatrix}
-aCov(X,X) + bCov(X,Y) + cCov(X,Z)\\
-dCov(Y,X) + eCov(Y,Y) + fCov(Y,Z)\\
-gCov(Z,X) + hCov(Z,Y) + iCov(Z,Z)
\end{bmatrix}
\end{equation}

Therefore,
\begin{multline}
Cov(D,D^{p})
=
sum \left(
\left(
\begin{bmatrix}
Cov(X,X) & Cov(X,Y) & Cov(X,Z) & 0\\
Cov(Y,X) & Cov(Y,Y) & Cov(Y,Z) & 0\\
Cov(Z,X) & Cov(Z,Y) & Cov(Z,Z) & 0\\
0 & 0 & 0 & 1
\end{bmatrix}
\right.
\right.
\times
\\
\left.
\left.
\begin{bmatrix}
-1 & 0 & 0 & 0\\
0 & 1 & 0 & 0\\
0 & 0 & 1 & 0\\
0 & 0 & 0 & 1
\end{bmatrix}
\bullet
\begin{bmatrix}
a & b & c & 0\\
d & e & f & 0\\
g & h & i & 0\\
0 & 0 & 0 & 1
\end{bmatrix}
\right)^T
\right)^T
\end{multline}

or
\begin{equation}
Cov(D,D^{p})=sum((Cov(D)\times RF \bullet RT)^T)^T
\label{covdd}
\end{equation}

Let $\ A $ be the dataset of attributes $\ (A_1, A_2, A_3,\dots, A_n) $, and let $\ V $ be the variance-covariance of a size $\ n\times n $ square matrix defined by,

\begin{equation}
	V_{i,j}=Cov(A_i,A_j)
\end{equation} 
where the diagonal elements are $\ V_{i,i}=Var(A_i)$

Let$\ r $ be a row vector of weight factors (in this case $\ n $ elements).
Then
\begin{equation}
Var(rX)=rVr^{'}
\end{equation}

written out in component form,

\begin{equation}
Var(r_1A_1+r_2A_2+r_3A_3+\dots+r_iA_i)=r_iCov(A_i,A_j)r_j 
\end{equation}

Since, $\ 	Cov(A_i,A_j)=Cov(A_i,A_j)$ and $\ Cov(A_i,A_i)=Var(A_i)$

\begin{equation}
r_{i}Cov(A_{i},A_{j})r_{j}=\forall i=j: r_{i}^{2}Var(A_i) + \forall i<j: 2 r_{i}r_{j}Cov(A_i,A_j)
\label{covmat}
\end{equation}

Hence, 
\begin{multline}
 Var(aX+bY+cZ)=a^{2}Var(X)+b^{2}Var(Y)+c^{2}Var(Z)+
 \\2abCov(X,Y)+2acCov(X,Z)+ 2bcCov(Y,Z)
\label{coveqprev} 
\end{multline}

Therefore,

\begin{multline}
Var(-aX+bY+cZ+\delta E)=a^{2}Var(X)+b^{2}Var(Y)+c^{2}Var(Z)-2abCov(X,Y)-
\\2acCov(X,Z)+ 2bcCov(Y,Z)
\label{coveq} 
\end{multline}
Hence,
\begin{multline}
Var(X^p)=a^{2}Var(X)+b^{2}Var(Y)+c^{2}Var(Z)-
\\2abCov(X,Y)-2acCov(X,Z)+ 2bcCov(Y,Z)
\label{varxp} 
\end{multline}
Similarly,\\
$Var(Y^p)=d^{2}Var(X)+e^{2}Var(Y)+f^{2}Var(Z)-2deCov(X,Y)-2dfCov(X,Z)+ 2efCov(Y,Z)$
$Var(Z^p)=g^{2}Var(X)+h^{2}Var(Y)+i^{2}Var(Z)-2ghCov(X,Y)-2giCov(X,Z)+ 2{ih}Cov(Y,Z)$

Hence, 

\begin{multline}
Var(D)
=diag \left(
\begin{bmatrix}
a & b & c & 0\\
d & e & f & 0\\
g & h & i & 0\\
0 & 0 & 0 & 1
\end{bmatrix}
\times
\begin{bmatrix}
-1 & 0 & 0 & 0\\
0 & 1 & 0 & 0\\
0 & 0 & 1 & 0\\
0 & 0 & 0 & 1
\end{bmatrix}
\times
\right.
\\
\left.
\begin{bmatrix}
Cov(X,X) & Cov(X,Y) & Cov(X,Z) & 0\\
Cov(Y,X) & Cov(Y,Y) & Cov(Y,Z) & 0\\
Cov(Z,X) & Cov(Z,Y) & Cov(Z,Z) & 0\\
0 & 0 & 0 & 1
\end{bmatrix}
\times
\begin{bmatrix}
-1 & 0 & 0 & 0\\
0 & 1 & 0 & 0\\
0 & 0 & 1 & 0\\
0 & 0 & 0 & 1
\end{bmatrix}
\times
\begin{bmatrix}
a & d & g & 0\\
b & e & h & 0\\
c & f & i & 0\\
0 & 0 & 0 & 1
\end{bmatrix}
\right)
\end{multline}

\begin{equation}
 Var(D^p)=diag(RT\times RF\times Cov(D)\times RF \times RT^T)
 \label{varddash}
\end{equation}

From Equations \ref{covdd} and \ref{varddash} we can write,

\begin{equation}
	Var(D-D^p)= \vv 1+diag(RT\times RF\times Cov(D)\times RF\times RT^T)+sum((Cov(D)\times RF \bullet RT)^T)^T
\end{equation}

Where, $ \vv 1 $ is a column vector of 1s of size $ n $, which is the number of attributes of $ D $. 
\par
Since, $\phi=min(Var(D-D^p))$,

\begin{multline}
	\phi= min(\vv1+diag(RT\times RF\times Cov(D)\times RF\times RT^T)+
	sum((Cov(D)\times RF \bullet RT)^T)^T)
\label{modifiedequation}
\end{multline}

\end{proof}
\endgroup
Now we can remove the two computationally intensive steps: $D^{p1}=(M_i\times TN^{noise}\times RF_{ax}\times (D^N)^T)^T$ and $\phi_{i,ax}=min(Var(x_{ij}-p_{ij})_{i=1}^m)_{j=1}^n$  $\forall$ $x_{ij}\in D^N$ with $\ \phi_{i,ax}=min(\vv1+diag(M_i\times RF_{ax}\times CD^N\times RF_{ax}\times M_i^T)+sum((CD^N\times RF_{ax} \bullet M_i)^T)^T)$.

\subsection{Steps of PABIDOT after efficiency optimization}

Algorithm \ref{parallelalgo} shows the  set of steps of the efficiency-optimized version of PABIDOT. In subsequent sections of the paper, PABIDOT (Algorithm \ref{parallelalgo}) refers to the optimized version of PABIDOT\_basic (Algorithm \ref{exhuativeapproach}). 

\begin{algorithm}[H]
	\caption{PABIDOT with efficiency optimization}\label{parallelalgo}
	\begin{spacing}{1.2}
		\begin{algorithmic}[1]
			\linespread{1}
			
			\LStatex\textbf{Inputs} :
			
			\begin{tabular}{l c l} 
				$D$  & $\gets $ & original dataset\\
                $\sigma$ & $\gets $ & input noise standard deviation (default value=0.3)\\
			\end{tabular}
			
			\LStatex \textbf{Outputs}:
			
			\begin{tabular}{ l c l } 
				$D^{p}$  & $\ \gets $ & perturbed dataset  \\
			\end{tabular}
			\LState $\Phi=0$
			\LState $\theta_{optimal} = 0$ 
			\LState $Rif_{optimal}=0$ 
			\LState generate $D^{N} $
            \LState generate $Cov(D^{N})$
			\LState generate $TN^{noise} $
			\For{\textbf{each} $ax$ in  $\left\{1,2,\dots,n\right\}$ } \label{al3step7}
			\LState generate $RF_{ax} $ 
			\For{\textbf{each} $\theta_i $ }
			\LState generate $M_i $ using Algorithm \ref{rotmatgenerate}
            \LState   $\ \phi_{i,{ax}}=min(\vv1+diag(M_i\times RF_{ax}\times Cov(D^{N})\times RF_{ax}\times M_i^T)+sum((Cov(D^{N})\times RF_{ax} \bullet M_i)^T)^T)$ \label{al2step11}	
			\EndFor			
			\EndFor \label{al3step18}
			\For{\textbf{each} $\theta_i $ }
            \LState   $\ \phi_{i}=min(\phi_{i,{ax}})$  where, $ax$ $\in$  $\left\{1,2,\dots,n\right\}$	
			\EndFor			
            \LState $\Phi = max(\phi_{i})$ where, $i$ $\in$  $\left\{1,2,\dots,\theta \right\}$
            \LState $\theta_{optimal}=\theta_i$  at $\Phi $
			\LState $Rif_{optimal}=ax$  at $\Phi $
            
			\LState generate $M\theta $ 
			\LState generate $RF_{optimal} $ 
			\LState$D^{pt}=(M\theta \times TN^{noise}\times RF_{optimal}\times (D^N)^T)^T$ \label{al3step21}
            \State $D^{pt}=(\left\| D^{pt}\right\|+\left\| \mathcal{N}(0,\sigma)\right\|$)$\bullet S_{\pm}$ \label{al2step22}
			\LState $D^{p} $=$\ D^{pt} \bullet STDVEC$+$MEANVEC$
			\State randomly swap the tuples of $\ D^{p} $ \label{al3swapstep} \label{al2step24}
			
			\Statex \textbf{End Algorithm}
		\end{algorithmic}
	\end{spacing}
\end{algorithm}
 
With the new derivation for $\phi$, the time consumption of PABIDOT drastically reduces. This can be observed in Figures \ref{timecomparisoninstances_newmethod} and \ref{timecomparison_bigoh} through the red colored plots. As shown, the efficiency optimization of PABIDOT (Algorithm \ref{parallelalgo}) introduces a considerable efficiency improvement over PABIDOT\_basic (Algorithm \ref{exhuativeapproach}). 
When the time consumption of PABIDOT and PABIDOT\_basic are plotted in the same diagram (Figure \ref{timecomparisoninstances_newmethod}), PABIDOT shows an almost constant trend. 
This is because of the increase in time consumption by PABIDOT after efficiency optimization is considerably lower than that of PABIDOT\_basic. However, the time consumption of PABIDOT is not constant, as shown in  Figure \ref{pabidot_timecomparisoninstances}.
Table \ref{scalability} shows the massive improvement of efficiency of PABIDOT after its efficiency optimization.  
 
\section{Experiments, Results, and Analysis}
\label{experiments}
This section provides information about the experimental setup, the results, and the analysis of the experimental results of PABIDOT's performance tests. The default input value of $\sigma=0.3$ is considered for the experiments unless specified otherwise. 
\subsection{Experimental Setup}
\label{expsetup}
\paragraph*{\textbf{Computation}} All the experiments except those on scalability and empirical privacy guarantee were tested on a Windows 7 (Enterprise 64-bit, Build 7601) computer with an Intel(R) i7-4790 (4$^{th}$ generation) CPU (8 cores, 3.60 GHz) and 8GB RAM. The scalability and the empirical privacy guarantee of the proposed algorithm were tested using a Linux (SUSE Enterprise Server 11 SP4) SGI UV3000 supercomputer, with 64 Intel Haswell 10-core processors, 25MB cache and 8TB of global shared memory connected by SGI's NUMAlink interconnect. 

 \paragraph*{\textbf{Datasets}} The datasets used for performance testing have different dimensions, i.e. varying from small to large, and contain only numerical attributes apart from the class attribute.  We removed the attributes that directly reflect the distribution of the class attribute to avoid classification accuracy bias. A short description of the nine datasets used for testing is given in Table \ref{datasettb}. Our dataset selection procedure was focused on selecting the datasets from a diverse range of domains (e.g. monetary data, wine quality data, document layout data, image features, binarized  regression data, electricity data, space shuttle data (NASA), exotic particle data, and signal process data) covering a diverse range of dimensions (e.g. 440$\times$8, 4898$\times$12, 5473$\times$11, 20000$\times$17, 40768$\times$11, 45312$\times$9, 58000$\times$9, 3310816$\times$28, and 11000000$\times$28). The selection of the datasets over other similar datasets was random and unbiased.

\pagebreak

\begin{table}[H]
\centering

    \caption{Short descriptions of the datasets selected for testing. We retrieved the datasets from the UCI ML data repository (https://archive.ics.uci.edu/ml/datasets/) and the OpenML data repository (https://www.openml.org/d/). The selected datasets' sizes range from small to extremely large, to allow the investigation of PABIDOT's performance on varying dimensions of data.}   
    \label{datasettb}

    \begin{small}
    	\setlength\tabcolsep{5pt} 
        \resizebox{1\columnwidth}{!}{
    \begin{tabular}{l l l l l }
    \hline
{\bfseries Dataset} & {\bfseries  Abbreviation}         & {\bfseries Number of Records}     & {\bfseries Number of Attributes }   &   \bfseries{Number of Classes}   \\
    \hline
Wholesale customers\tablefootnote{https://archive.ics.uci.edu/ml/datasets/Wholesale+customers}       &	 WCDS & 440 \ & 8 \ & 2 	\\
    Wine Quality\tablefootnote{https://archive.ics.uci.edu/ml/datasets/Wine+Quality}      & WQDS & 4898 \ & 12 \ & 7 \\
     Page Blocks Classification \tablefootnote{https://archive.ics.uci.edu/ml/datasets/Page+Blocks+Classification}          & PBDS  & 5473 \ &  11 \ &   5  \\
Letter Recognition\tablefootnote{https://archive.ics.uci.edu/ml/datasets/Letter+Recognition}             &  LRDS	& 20000 & 17 & 26\\
Fried\tablefootnote{https://www.openml.org/d/901}      & FRDS & 40768  \ & 11 \ & 2 \\
Electricity\tablefootnote{https://www.openml.org/d/151}      & ELDS & 45312  \ & 9 \ & 2 \\
       Statlog (Shuttle)\tablefootnote{https://archive.ics.uci.edu/ml/datasets/Statlog+\%28Shuttle\%29}        &  SSDS & 58000 \ & 9 \ & 7 \\

HEPMASS\tablefootnote{https://archive.ics.uci.edu/ml/datasets/HEPMASS\#}      & HPDS & 3310816  \ & 28 \ & 2 \\
HIGGS\tablefootnote{https://archive.ics.uci.edu/ml/datasets/HIGGS\#}      & HIDS & 11000000  \ & 28 \ & 2 \\

    \hline
    \end{tabular}
    }
    \end{small} 
    
\end{table}

\paragraph*{\textbf{Classification algorithms}}\label{classificationalgo} Data classification tests were carried out by using Weka 3.6, ~\cite{witten2016data} a collection of machine learning algorithms for data mining tasks. As different classes of classification algorithms employ different classification strategies; we used five classification algorithms for testing the accuracy of the proposed method: Multilayer perceptron (MLP), k-nearest neighbor (kNN), Sequential Minimal Optimization (SMO), Naive Bayes, and J48 ~\cite{witten2016data}. MLP is a  classifier that uses back-propagation to classify instances, kNN  is a non-parametric method used for classification ~\cite{witten2016data}. SMO is an implementation of John Platt's sequential minimal optimization algorithm for training a support vector classifier. Naive Bayes is a fast classification algorithm based on probabilistic classifiers. J48 is an implementation of the decision tree classification algorithm ~\cite{witten2016data}.

\subsection{Performance Evaluation}
Performance evaluation focuses on the utility which can be defined as the usability or the effectiveness of a perturbation method and its underlying data. We investigated the utility of PABIDOT in terms of classification accuracy.  However, PABIDOT must prove its usability based on time complexity, memory overhead, scalability and biases, which formed the foundations to our evaluation. We compared the results of PABIDOT with the results obtained from random rotation perturbation (RP) ~\cite{chen2005random} and geometric perturbation (GP) ~\cite{chen2011geometric} as they are the closest methods to PABIDOT. 

The original datasets were perturbed using  RP, GP, and PABIDOT, and the classification accuracy generated by the perturbed data of the three methods were compared. The comparisons were made using a nonparametric statistical comparison test: Friedman's rank test, which is analogous to a standard one-way repeated-measures analysis of variance ~\cite{howell2016fundamental}. Friedman's mean ranks (FMR) and the statistical significance of the results were recorded. We tested the time complexity of PABIDOT via theoretical and runtime measurements. The scalability of PABIDOT was tested by running the algorithm on the SGI UV3000 supercomputer over the large-scale datasets, HPDS and HIDS then comparing the runtime results to that of RP and GP. Finally, the memory requirements of the three methods were investigated by looking at peak memory usage; for that, MATLAB's memory profiler tool~\cite{nell2011memory} was used.

\subsubsection{Time Complexity} Algorithm \ref{parallelalgo} has two loops that are controlled by the number of attributes in the dataset and by the rotation angle range. Since the bounds of the rotation angle are fixed, the internal loop is repeated a constant number of times, and we denote this constant as  $k$. For a dataset of $n$ number of attributes and $m$ number of tuples, the outer loop iterates for $n$ number of times. The estimated time complexity of the loops is then $O(k\times n)=O(n)$.  By taking into consideration that step \ref{al2step11} is $O(n^3)$, loop block (from step \ref{al3step7} to \ref{al3step18}) has a complexity of $O(n^4)$. It can be identified that the most computationally complex step out of the remaining steps is step \ref{al3step21}, which has a complexity of $O(n^3\times m)$. Therefore, we can conclude that the worst case complexity of PABIDOT is governed by  $O(n^3\times m)$ as $m>>>n$.

\begin{figure}[H] 
	
	\centering
	\subfloat[Time consumption of PABIDOT against the number of tuples.]{\includegraphics[width=0.48\textwidth, trim=0cm 0cm 0cm 0cm]{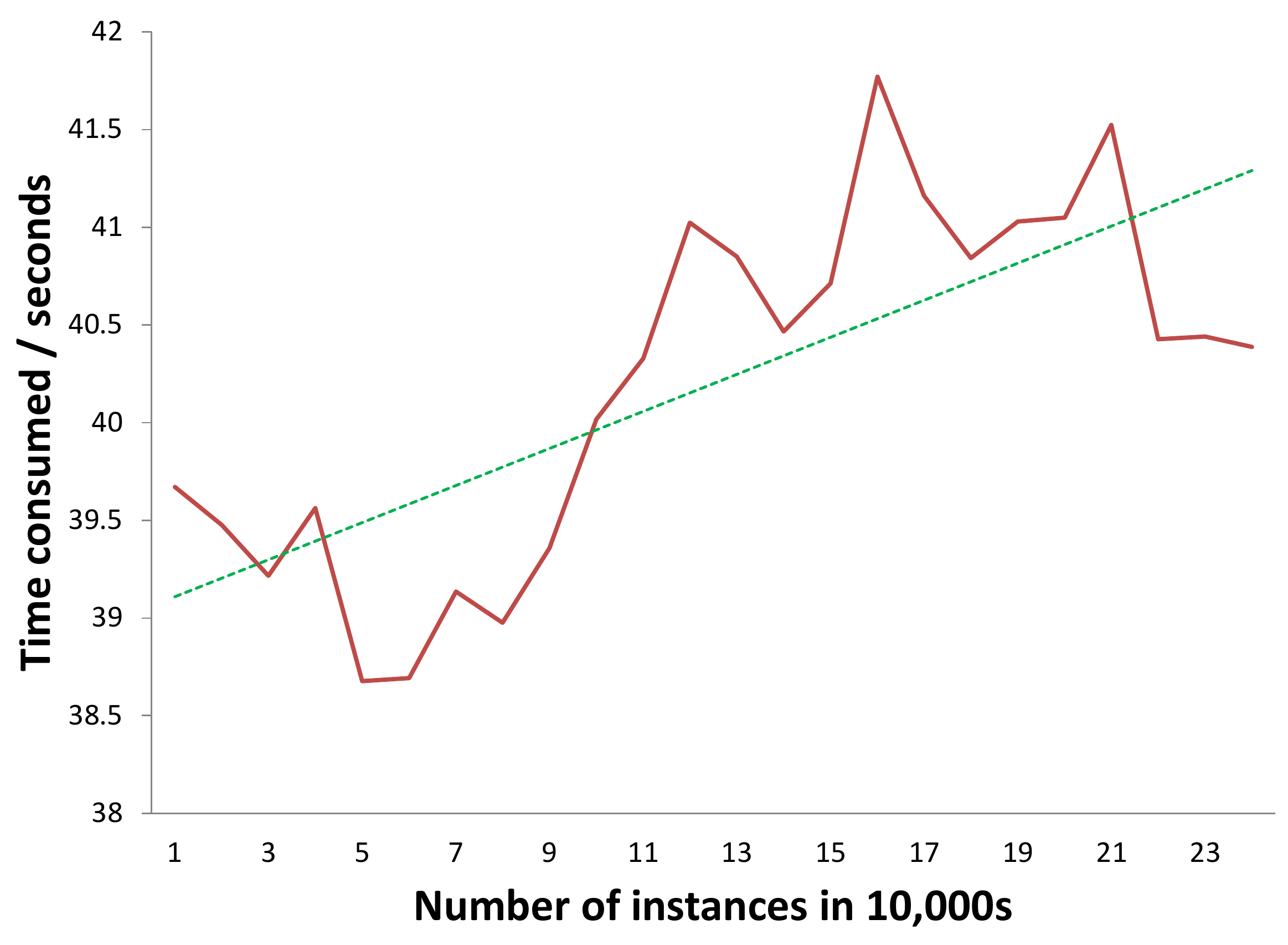}\label{pabidot_timecomparisoninstances}}
	\hfill
	\subfloat[Time consumption of PABIDOT against the number of attributes.]{\includegraphics[width=0.48\textwidth, trim=0cm 0cm 0cm 0cm]{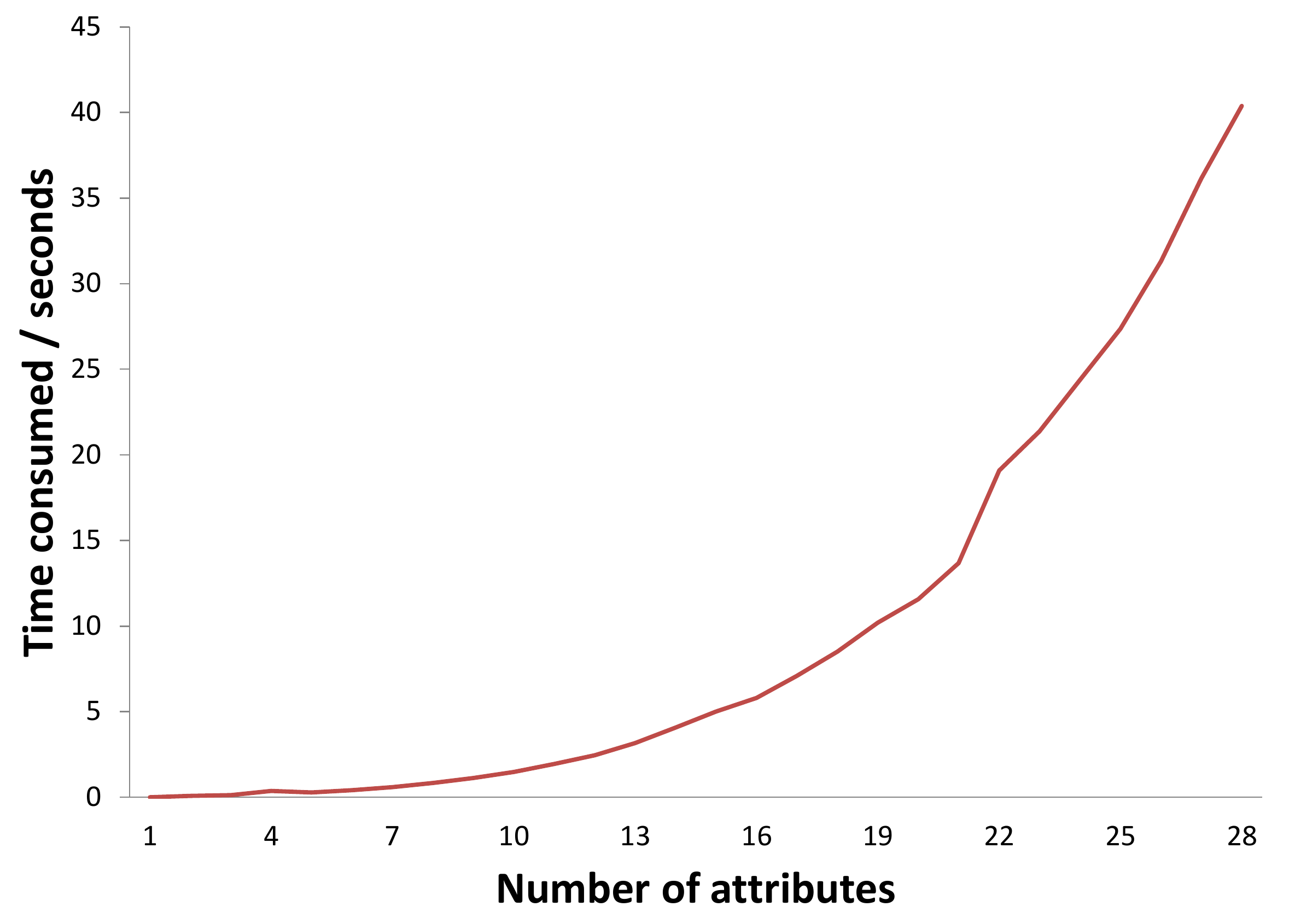}\label{pabidot_timecomparison}}    
	\caption{Time consumption of PABIDOT. PABIDOT shows linear time complexity for the number of instances, and exponential time complexity for the number of attributes. However, the time consumption is very low in all cases.}
    \label{timecompthree}

\end{figure}

\begin{figure}[H] 
	\centering
	\subfloat[Time consumption of PABIDOT against the number of tuples before and after the efficiency optimization.]{\includegraphics[width=0.48\textwidth, trim=0cm 0cm 0cm 0cm]{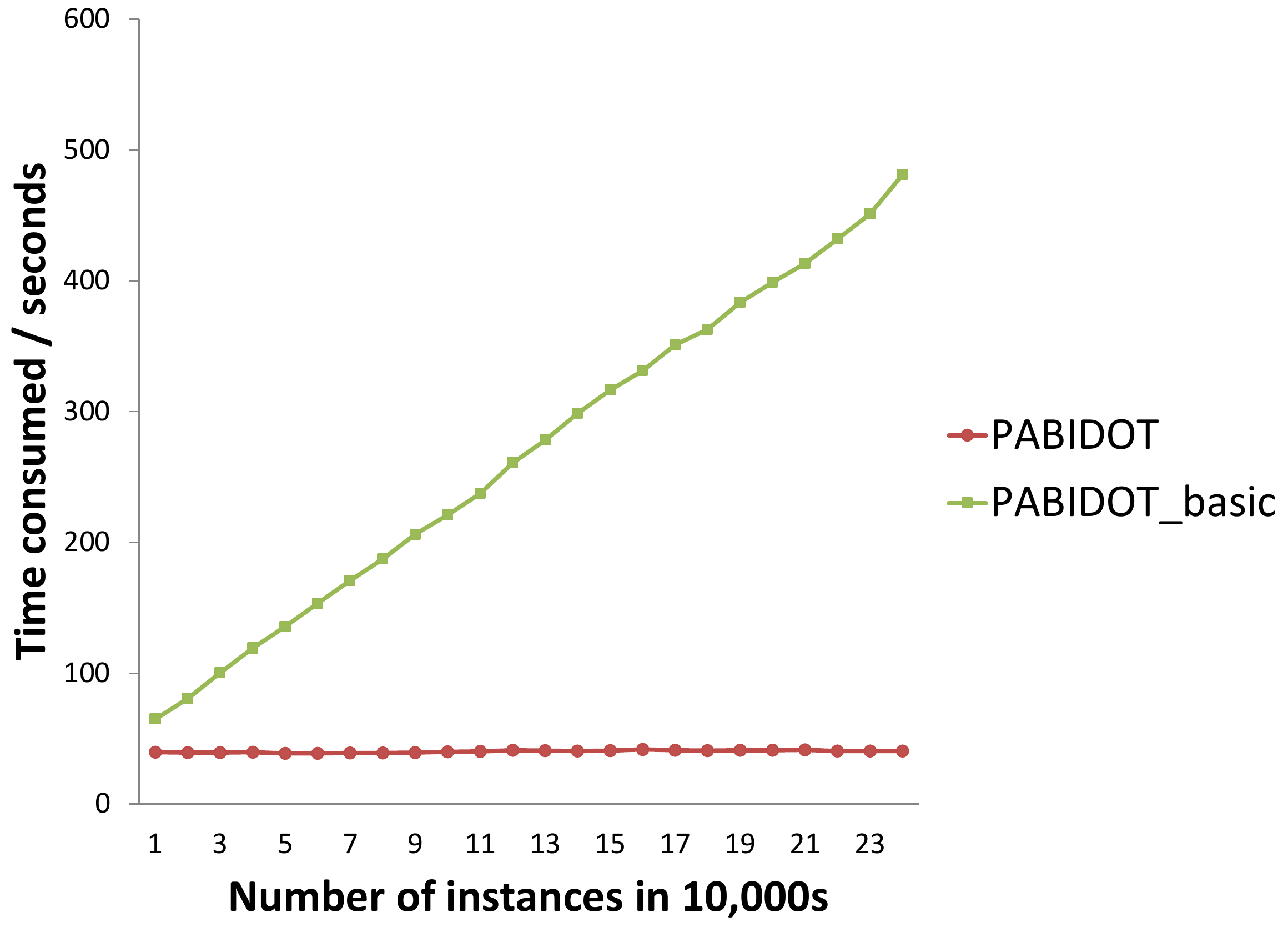}\label{timecomparisoninstances_newmethod}}
	\hfill
	\subfloat[Time consumption of PABIDOT against the number of attributes before and after the efficiency optimization.]{\includegraphics[width=0.48\textwidth, trim=0.3cm 0cm 0cm 0cm]{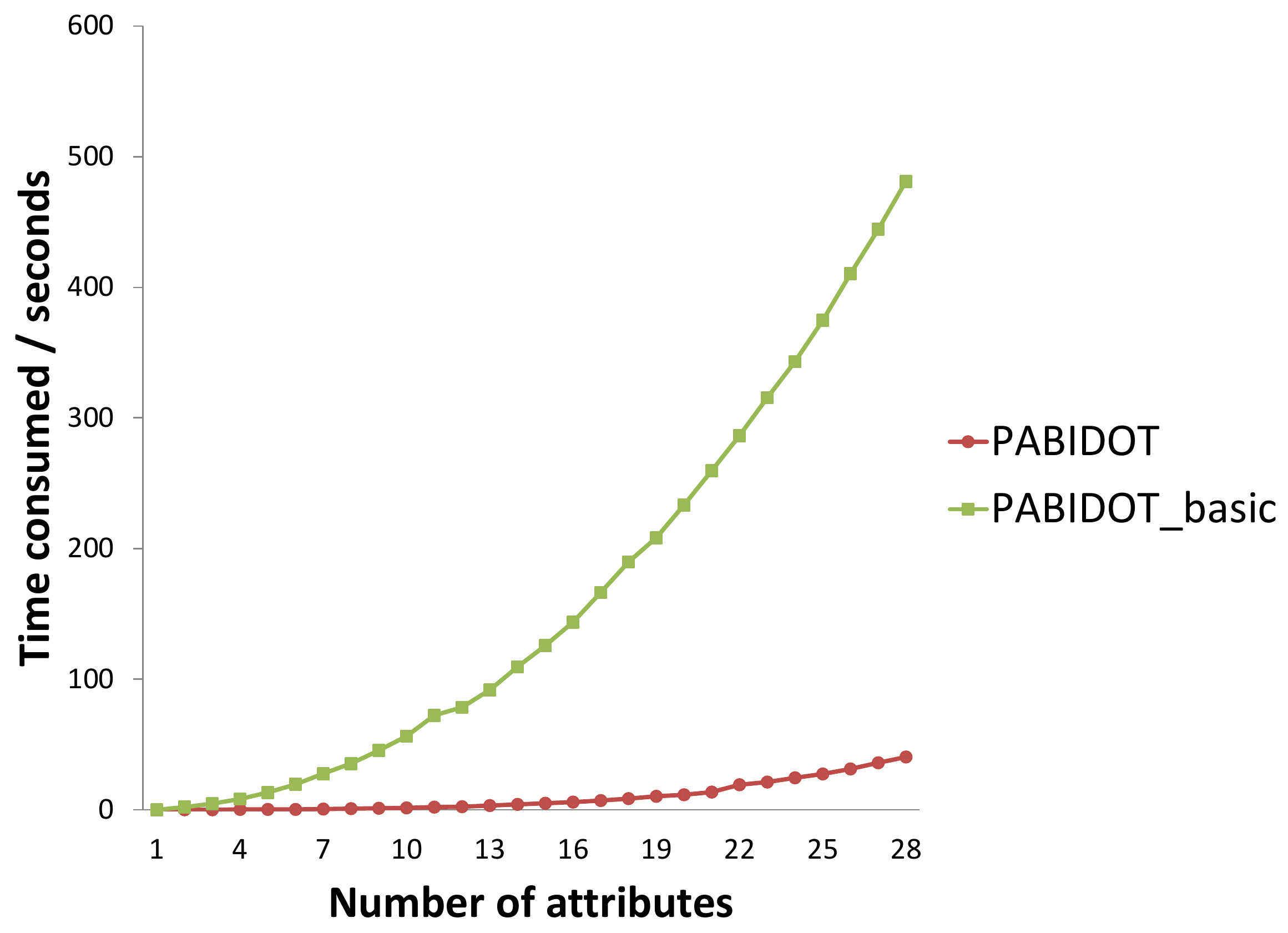}\label{timecomparison_bigoh}}

	\caption{Time Consumption of PABIDOT before and after the efficiency optimization. Both PABIDOT and PABIDOT\_basic show similar trends in time consumption in the respective categories; however, PABIDOT is faster.}
    \label{twotrends}

\end{figure}

\paragraph*{\textbf{Effect of the number of instances(tuples) and the number of attributes on the perturbation time taken by PABIDOT}} We used the HPDS dataset to obtain a clear idea about the trends in time consumption of PABIDOT. The dataset was trimmed (t\_HPDS) to have the dimensions of 240550$\times$28 which is large enough to provide a significant trend while keeping the execution time within reasonable bounds. Figure \ref{pabidot_timecomparisoninstances} shows the time consumption of PABIDOT for increasing number of instances while the number of attributes is constant. Figure \ref{pabidot_timecomparison} shows the time consumption of PABIDOT for an increasing number of attributes while the number of instances is constant. 
In Figure \ref{twotrends} the same two plots were drawn together with the corresponding times before efficiency optimization (i.e. PABIDOT and PABIDOT\_basic together).

\begin{figure}[H] 
	
	\centering
	\subfloat[Increase of time consumption of the three methods against the number of tuples.]{\includegraphics[width=0.48\textwidth, trim=0cm 0cm 0cm 0cm]{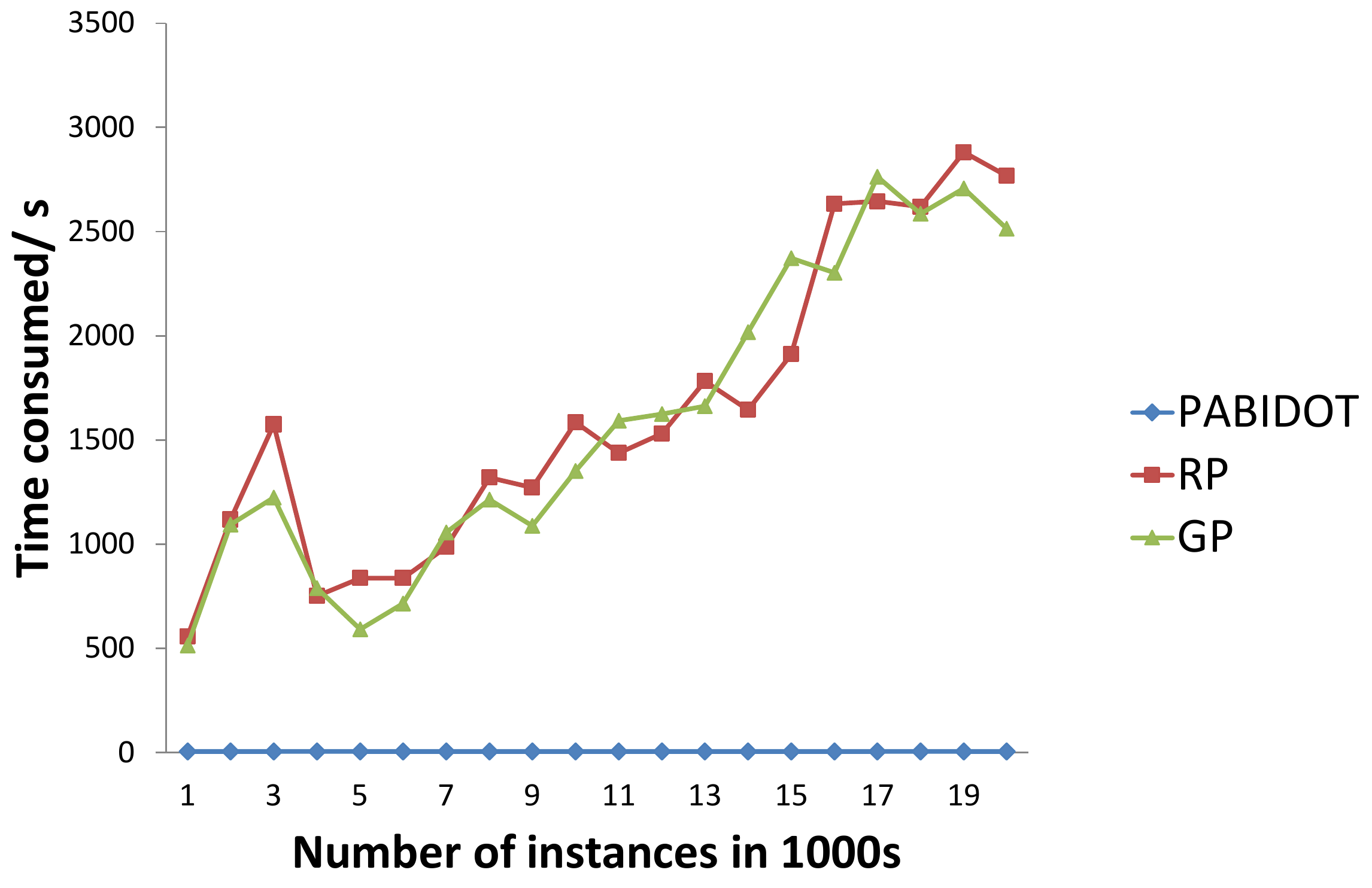}\label{timecomparisoninstances}}
	\hfill
	\subfloat[Time consumption of the three methods against the number of attributes.]{\includegraphics[width=0.48\textwidth, trim=0cm 0cm 0cm 0cm]{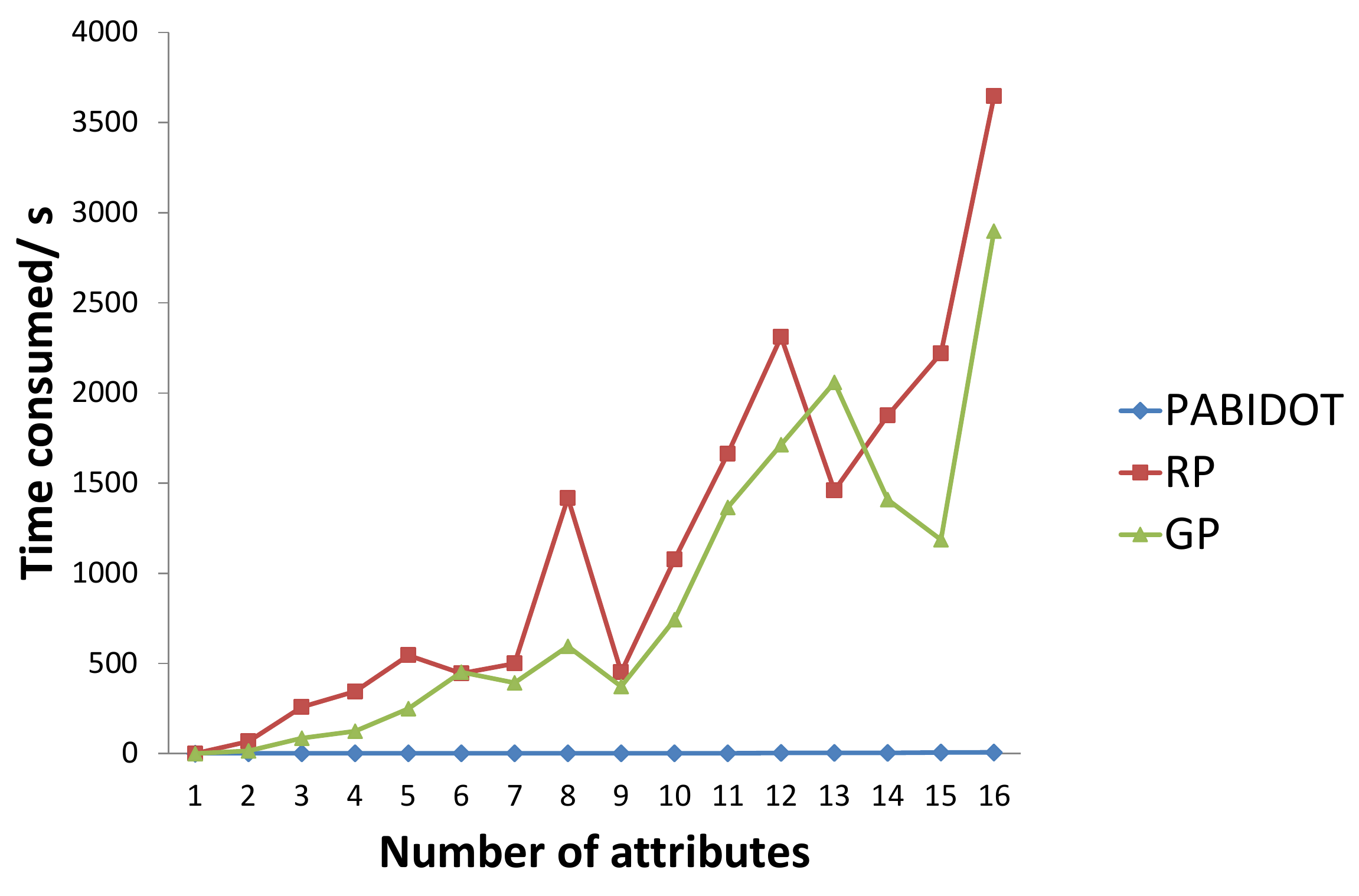}\label{timecomparison}}    
	\caption{Time consumption comparison of the three methods. Due to the extremely low time consumption of PABIDOT, its curve lies almost on the x-axis when drawn in a plot together with the others. The plots further show that the increase in the number of instances and the increase in the number of attributes have a very small impact on PABIDOT's time consumption, compared to RP and GP.}
	

    \label{pabidot_timecompthree}
 
\end{figure}

\paragraph*{\textbf{Effect of the number of instances and the number of attributes on perturbation time: comparison of RP, GP, and PABIDOT}} t\_HPDS was too large for RP and GP, therefore, LRDS was used to test and compare the time consumption of the three methods with increasing number of instances and increasing number of attributes respectively.  In Figure \ref{pabidot_timecompthree} it can be noticed that the time consumption of PABIDOT is extremely low compared to RP and GP.

\subsubsection{Scalability}
We conducted the scalability analysis on the SGI UV3000 supercomputer.  The results are provided in Table \ref{scalability}. The results for both before ($PABIDOT\_basic$) and after efficiency optimization of PABIDOT are provided along with the results of RP and GP. The plots compare the efficiency of PABIDOT compared to RP and GP.  The plots also show the efficiency improvement of PABIDOT as a result of optimization.

\begin{table}[H]
	\centering
	
	\caption{Scalability of the three algorithms for the high dimensional data.  We ran the corresponding runtime experiments on the SGI UV3000 supercomputer. The scripts were set to a time limit of 100 hours, but RP and GP couldn't finish the perturbation process for either of datasets even with this generous time limit.}     
	\label{scalability}
	\setlength\tabcolsep{5pt} 
	\begin{small}
		
        \resizebox{0.9\columnwidth}{!}{
		\begin{tabular}{ l l l l l l}
			\hline
			{\bfseries Dataset} & {\bfseries  Dimensions}           & {\bfseries RP  } & {\bfseries GP  } & {\bfseries PABIDOT\_basic } & {\bfseries PABIDOT} \\
		
			\hline
			HPDS        & 3310816$\times$28 & Not converged   & Not  converged 	& 2.9 hours & 71.41 seconds \\
			
			 &	  & for 100 hours & for 100 hours &\\
             HIDS        & 11000000$\times$28 & Not converged   & Not  converged 	& 11.16 hours &  100.45 seconds\\
			
			 &	  & for 100 hours & for 100 hours &\\

			\hline
		\end{tabular}
        }
	\end{small} 
	

\end{table}

\subsubsection{Memory Overhead}
The peak memory consumptions of the three methods for the first seven datasets (available in Table \ref{datasettb})  are provided in Table \ref{memoryconsumption}. The relevant Friedman test ranks are provided in the last row of the table.  The experiment returned the test statistics of a $\chi^2$ value of 3.7143, a degree of freedom of 2 and a p-value of 0.1561. The p-value says that the memory consumptions of the three methods have more or less a similar pattern. However, FMR values suggest that PABIDOT consumes the lowest amount of memory among the three methods.

\begin{table}[H]
\centering

    \caption{Peak memory (Kb) consumption data of the three methods for the first seven datasets in Table \ref{datasettb}. The peak memory consumptions recorded the highest memory usage of the corresponding methods at a particular instance in kilobytes. We used Matlab memory profiler to track the peak memory consumption during the perturbation of each dataset.}     
    \label{memoryconsumption}

    \begin{small}
    \resizebox{0.4\columnwidth}{!}{
    \begin{tabular}{l l l l}
    \hline
{\bfseries Dataset}       & {\bfseries RP }     & {\bfseries GP  } & {\bfseries  PABIDOT } \\
    \hline
    LRDS    	&  2564 & 3084 & 2664\\
    PBDS      &  1540 &  2056 & 428\\
    SSDS    	& 3072	& 3180 & 3408\\
    WCDS    	&  512 & 1024  & 444\\   
    WQDS    	& 512	& 5846  & 184\\
    
    FRDS    	&  4756 &  3853  & 3865\\   
    ELDS    	&  4945	&  6312  & 3449 \\
    \hline
    
    \hline
    {\bfseries FMR} & 1.86  &  2.57  & 1.57 \\
    \hline
    \end{tabular}
    }
    \end{small} 
    

\end{table}

\subsubsection{Classification Accuracy}

\begin{figure}[H]
	\centering
	\scalebox{1}{
	\includegraphics[width=1\textwidth, trim=0cm 0cm 0cm 0cm]{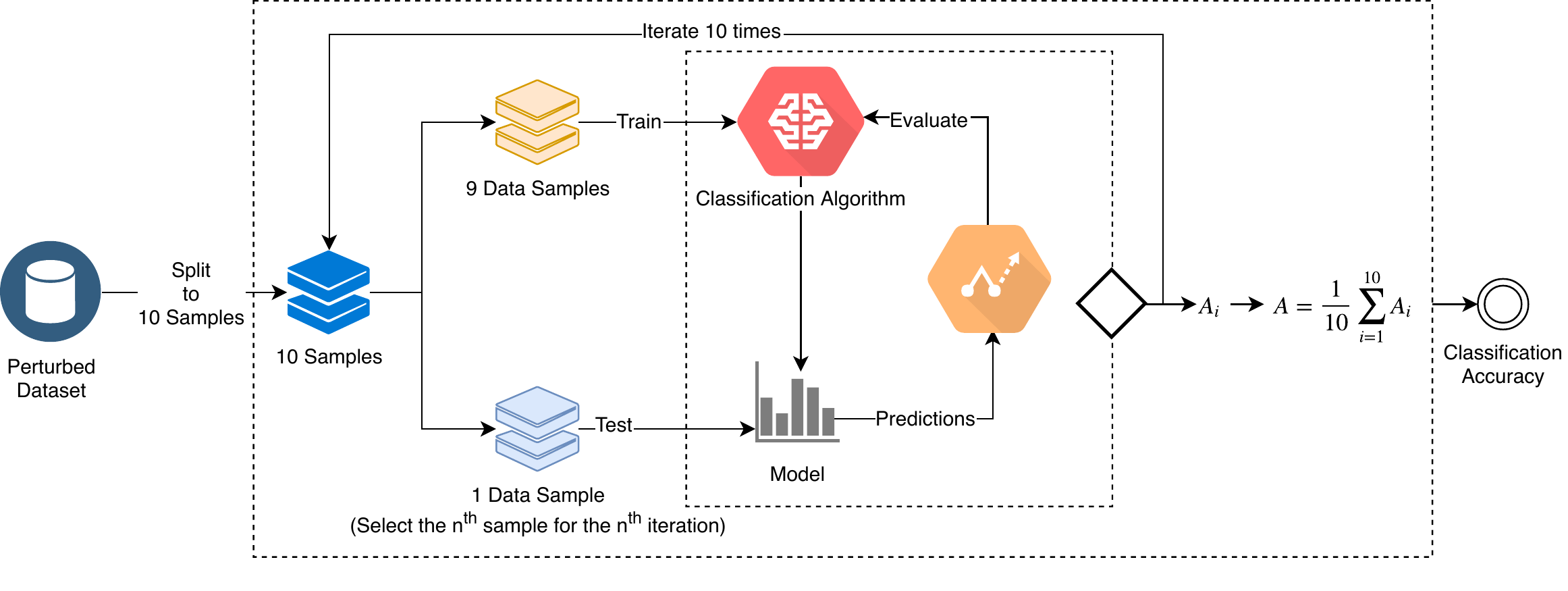}
	}
	\caption{The process used to generate the classification models trained by the perturbed data. This figure represents the overall process of 10-fold cross-validation used in our experiments. We validated the classification models for each of the perturbed datasets. Ten samples were taken from each particular perturbed dataset, and the validation ran for ten iterations. In a single iteration, nine samples out of 10 were used for training, and 1 sample for testing. Finally, the average classification accuracy of the ten cycles was returned as the final accuracy for a particular dataset.}
	

	\label{clssiflowchart}
\end{figure}

Table \ref{classyaccuracy} has the classification accuracy results of the original dataset and the datasets perturbed by RP, GP, and PABIDOT. 10-fold cross-validation was used to validate the classification models. An instance of the model validation for a perturbed dataset is presented in Figure \ref{clssiflowchart}. As shown in Figure \ref{clssiflowchart}, for the case of perturbed data (RP, GP, and PABIDOT), both classification model training and testing are done using the same perturbed dataset.

\begin{figure}[H]
	\centering
	\scalebox{0.8}{
	\includegraphics[width=1\textwidth, trim=0cm 0cm 0cm 0cm]{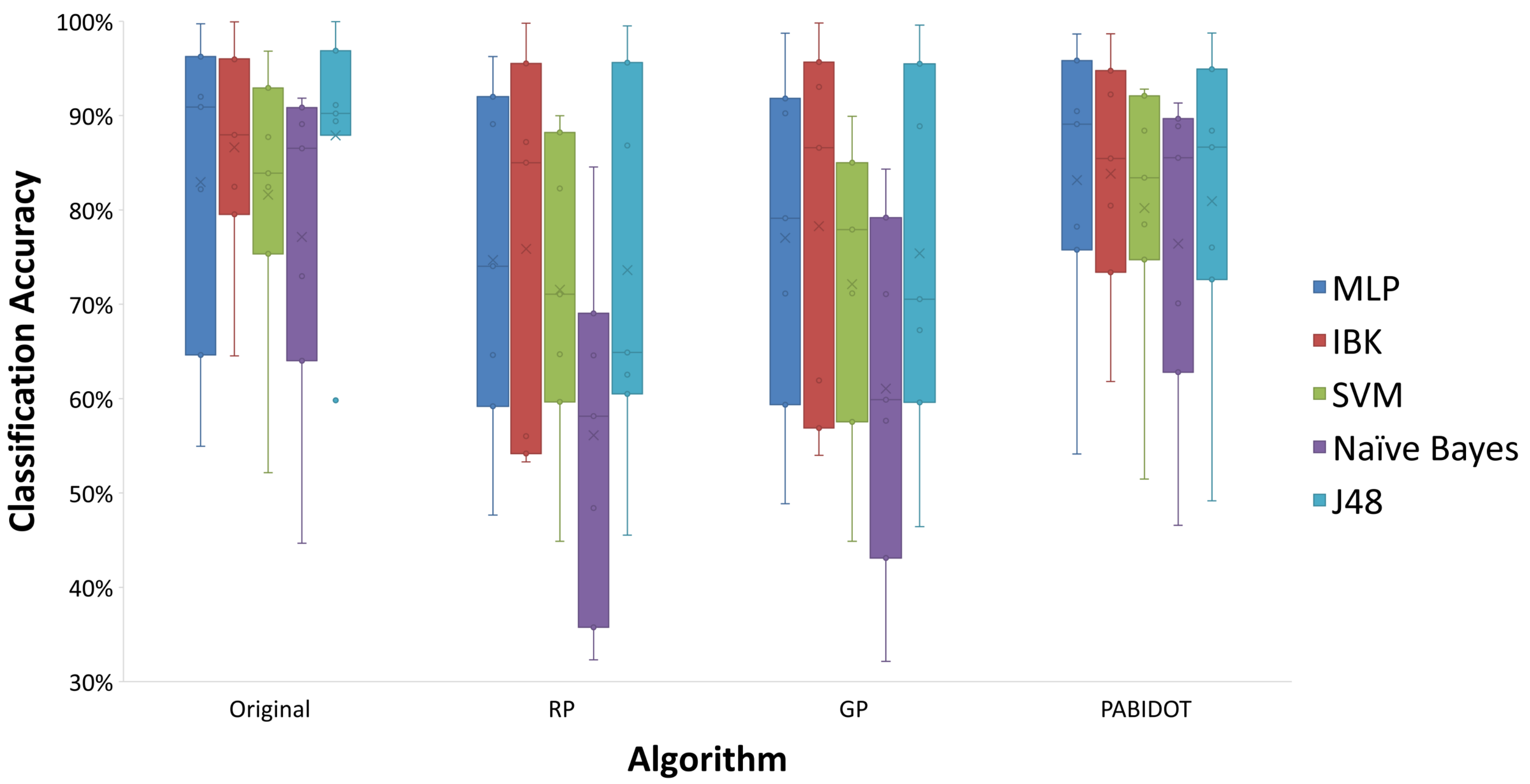}
	}
	\caption{Box plots for the datasets listed in Table \ref{classyaccuracy}. The boxplots in the figure show how each perturbation algorithm performs on different classification algorithms. The box plots belonging to PABIDOT show higher medians, $25^{th}$ percentiles, and $75^{th}$ percentiles compared to RP and GP suggesting that PABIDOT produces better utility in terms of classification accuracy compared to RP and GP.}
	\label{accuracy_comparison}


\end{figure}

The percentage values in Table \ref{classyaccuracy} are the average accuracy values returned for the first seven datasets shown in Table \ref{datasettb}.  The classification accuracy results were compared using Friedman's rank test. The last row of Table \ref{classyaccuracy} has Friedman's ranks for the classification accuracies of the three methods. The test statistics of the experiment had a $\chi^2$ value of 24.7914, a degree of freedom of 2 and a p-value of 4.1364e-06. Figure \ref{accuracy_comparison} shows the box plots for the datasets listed in Table \ref{classyaccuracy}.

\begin{table}[htbp]
  \caption{Classification accuracies returned by the original dataset and the three methods. The last row shows the Friedman's mean ranks (FMRs) of the classification accuracy data which correspond to each of the methods (RP, GP, and PABIDOT). A higher rank indicates that the corresponding method tends to provide higher classification accuracy. Consequently, PABIDOT  is likely to return higher accuracies compared to RP and GP. }  
   \label{classyaccuracy}
    \centering 
    \resizebox{0.8\columnwidth}{!}{
    \begin{tabular}{rllllll}
    \toprule
    \multicolumn{1}{l}{\textbf{Dataset}} & \textbf{Algorithm} & \multicolumn{1}{l}{\textbf{MLP}} & \multicolumn{1}{l}{\textbf{IBK}} & \multicolumn{1}{l}{\textbf{SVM}} & \multicolumn{1}{l}{\textbf{Naive Bayes}} & \multicolumn{1}{l}{\textbf{J48}} \\
    \midrule
    \multicolumn{1}{l}{LRDS} & Original   & 82.20\% & 95.96\% & 82.44\% & 64.01\% & 87.92\% \\
          & RP    & 74.04\% & 87.19\% & 71.07\% & 48.41\% & 64.89\% \\
          & GP    & 79.12\% & 93.05\% & 77.92\% & 59.89\% & 70.54\% \\
          & PABIDOT & 78.22\% &	92.24\%	& 78.48\% &	62.80\% &	72.62\%\\
    \multicolumn{1}{l}{PBDS} & Original   & 96.25\% & 96.02\% & 92.93\% & 90.85\% & 96.88\% \\
          & RP    & 92.00\% & 95.52\% & 89.99\% & 35.76\% & 95.61\% \\
          & GP    & 90.24\% & 95.67\% & 89.93\% & 43.10\% & 95.49\% \\
          & PABIDOT & 95.83\% &	94.76\% &	92.09\% &	89.68\% &	94.92\%\\
    \multicolumn{1}{l}{SSDS} & Original   & 99.72\% & 99.94\% & 96.83\% & 91.84\% & 99.96\% \\
          & RP    & 96.26\% & 99.80\% & 88.21\% & 69.04\% & 99.51\% \\
          & GP    & 98.73\% & 99.81\% & 78.41\% & 79.18\% & 99.59\% \\
          & PABIDOT & 98.65\% &	98.67\% &	92.80\% &	91.34\% &	98.74\%\\
    \multicolumn{1}{l}{WCDS} & Original   & 90.91\% & 87.95\% & 87.73\% & 89.09\% & 90.23\% \\
          & RP    & 89.09\% & 85.00\% & 82.27\% & 84.55\% & 86.82\% \\
          & GP    & 91.82\% & 86.59\% & 85.00\% & 84.32\% & 88.86\% \\
          & PABIDOT & 90.45\% &	85.45\% &	88.41\% &	88.86\% &	88.41\%\\
    \multicolumn{1}{l}{WQDS} & Original   & 54.94\% & 64.54\% & 52.14\% & 44.67\% & 59.82\% \\
          & RP    & 47.65\% & 53.29\% & 44.88\% & 32.32\% & 45.53\% \\
          & GP    & 48.86\% & 56.88\% & 44.88\% & 32.16\% & 46.43\% \\
          & PABIDOT & 54.12\% &	61.82\% &	51.47\% &	46.57\% &	49.16\%\\
    \multicolumn{1}{l}{FRDS} & Original   & 91.99\% & 82.46\% & 83.89\% & 86.53\% & 89.41\% \\
          & RP    & 64.62\% & 56.03\% & 64.71\% &  64.59\% & 62.53\% \\
          & GP    & 71.14\% & 61.92\% & 71.16\% & 71.08\% & 67.25\% \\
          & PABIDOT & 89.09\% &	80.45\% &	 83.41\% &	85.53\% &	86.65\%\\
    \multicolumn{1}{l}{ELDS} & Original   & 64.63\% & 79.53\% & 75.34\% & 72.98\% & 91.11\% \\
          & RP    & 59.18\% & 54.18\% & 59.65\% & 58.14\% & 60.50\% \\
          & GP    & 59.36\% & 53.99\% & 57.54\% & 57.65\% & 59.60\% \\
          & PABIDOT & 75.77\% &	73.39\% &	74.73\% &	70.10\% &	76.03\%\\
          \midrule
          \multicolumn{1}{c}{\bfseries FMR Values}& \multicolumn{2}{c}{RP: 1.41 }& \multicolumn{2}{c}{GP: 1.98 }& \multicolumn{2}{c}{PABIDOT: 2.60 }\\
    \bottomrule
    \end{tabular}%
    }
    

\end{table}%

\subsubsection{Biases and post-processing properties} \label{biases}
Perturbation and randomization have a noticeable impact on the summary statistics such as mean and standard deviations. Therefore, PABIDOT can be considered to have Type A biases. We ran a two-sample Kolmogorov-Smirnov test between the tuples of the original and perturbed datasets to check whether the perturbation has made an impact on the relationship between the confidential attributes. The test was carried out at 5\% significance level. Table \ref{kolmogorovtest} includes the number of tuples that accepted the null hypothesis indicating that the perturbed and non-perturbed tuples have the same continuous distribution.

\begin{table}[H]
	\caption{Two-sample Kolmogorov-Smirnov test for the tuple comparison of the perturbed and non-perturbed datasets. We ran two-sample Kolmogorov-Smirnov test on each of the associated (original$\leftrightarrow$perturbed) records of the original data and the data perturbed by PABIDOT. ``Total" shows the number of records that returned a test value of ``0" which indicates that two-sample Kolmogorov-Smirnov test does not reject the null hypothesis at the 5\% significance level, meaning that the corresponding records are similar. ``Percentage" shows the percentage of the total number of similar tuples compared to the total number of records in each dataset. According to the percentages, it is clear that PABIDOT maintains the record integrity of the perturbed data.}
    \centering 
    \resizebox{0.7\columnwidth}{!}{
		\begin{tabular}{l l l l l l l l} 
			\toprule
			\bfseries Dataset & \bfseries LRDS & \bfseries PBDS & \bfseries SSDS &  \bfseries WCDS & \bfseries WQDS & \bfseries FRDS & \bfseries ELDS\\
			\midrule
			\bfseries Total & 16090 & 1928 & 37629 & 434 & 4897 & 38222 & 43034\\ 
			\bfseries Percentage & 80.45\% & 35.23\% & 86.50\% & 98.64\% & 99.98\% & 93.75\% & 94.97\%\\ 
			\bottomrule
			\label{kolmogorovtest}
		\end{tabular}
        }
        

\end{table}

From Table \ref{kolmogorovtest}, it is evident that the new perturbation algorithm suffers a low level of Type B biases. The results under Type B bias can be extended to conclude that the perturbation suffers only a mild level of Type C biases, assuming that the perturbation process is conducted on both confidential and non-confidential attributes. Next, a two-sample Kolmogorov-Smirnov test was run between the attributes (perturbed vs. non-perturbed) of the datasets. The test was carried out at 5\% significance level. The null hypothesis was rejected with a probability of 100\% in almost all the cases proving that the perturbation process makes changes in the attribute distributions, which in turn proves that PABIDOT introduces Type D bias to the dataset.

\subsection{Privacy Analysis}

The risk of disclosure by a perturbation algorithm can be measured by quantifying the underlying privacy.  Three privacy aspects of PABIDOT were analyzed: attack resistance, privacy guarantee, and the increase in information entropy. PABIDOT was tested for known I/O attacks, and ICA-based attacks that are based on data reconstruction. PABIDOT's attack resistance results were compared with RP and GP using Friedman's rank test, and the results were presented with the corresponding test statistics. Further, attack resistance was analyzed employing induction using its characteristic properties of perturbation and the properties of the attacks such as complementary release attacks (Table \ref{risklevel}). PABIDOT perturbs a dataset at the best perturbation parameters to achieve $\Phi-separation$ privacy. This feature of PABIDOT was proven by testing the dynamics of $\Phi$ against data reconstruction attacks as summarized in Figure \ref{phi_attack_std}.

Increase in entropy is a useful metric to measure the impurity of a dataset \cite{park2003anomaly}.  Since, a large entropy value implies the availability of a large number of different records, the increase of entropy can be a good measurement to define the difficulty of data reconstruction ~\cite{park2003anomaly}.  Therefore, the perturbed data were tested for their impurity using the increase of entropy to establish a further estimation of the level of privacy provided by PABIDOT.

\subsubsection{Attack Resistance} \label{attacktesting}
The literature shows different attack types that are probable against matrix multiplicative data perturbation. The primary purpose of these attacks is to restore the original data from the perturbed data. Known I/O attacks ~\cite{okkalioglu2015survey}, known sample attacks ~\cite{okkalioglu2015survey} and ICA (Independent Component Analysis) ~\cite{okkalioglu2015survey} based attacks are three well-known attack types targeting matrix multiplicative data perturbation. Since PABIDOT is based on matrix multiplication,  it was initially tested against two typical attacks, namely known as I/O and ICA based attacks, and resistance to further attack methods are under consideration.  As the privacy metric used in the new method optimizes the level of privacy provided by the perturbation against naive estimation, the final perturbed dataset guarantees that it provides enough resistance to naive estimation.

\paragraph*{\textbf{Known I/O and ICA-based attacks}} To check the resistance of the proposed algorithm to ICA based attacks, the procedure described in ~\cite{chen2005random} was employed, and the FastICA package ~\cite{gavert2005fastica} was used to evaluate the effectiveness of ICA-based reconstruction of the perturbed data.  The attack resistance experiment was inspired by the experiments described in ~\cite{chen2005random,chen2011geometric}, and the results are presented in Table \ref{attackresistance}. These values were obtained as standard deviation of the difference between the normalized original data and the perturbed data (NI) and reconstructed data (reconstructed using ICA and IO). Our assumption was that 10\% of the original data is known to the adversary in the known I/O attack investigation. The results returned by RP and GP were generated using 10 number of iterations with a noise factor (sigma) of 0.3 (the default settings). The subscripts ``$avg$" and ``$min$" in the title row represent the minimum and average values of the corresponding data. The ``$min$" values under each test indicate the minimum guarantee of resistance, while ``$avg$" values give an impression of the overall resistance.

\begin{table}[H]
	\centering
	
	\caption{Attack resistance of the algorithms. The columns NI, ICA, IO represent naive inference, independent component analysis, and known input/output attacks respectively. NI examines the difference between the original and perturbed data.  ICA and IO attacks were run on each perturbed dataset to reconstruct datasets. In ICA and IO, the difference between the original and the reconstructed data was considered. $min$ and $avg$ represent the minimum and average of the standard deviation values of the difference between original and reconstructed data respectively. A higher value represents a higher difference, implying higher resistance. Friedman's mean rank (FMR) of PABIDOT indicates that it provides higher resistance compared to RP and GP.}     
	\label{attackresistance}
	
	\begin{small}
    \resizebox{0.9\columnwidth}{!}{
		\begin{tabular}{l l l l l l l l l }
			\hline
			{\bfseries Dataset} & {\bfseries Algorithm} &  {\bfseries  $NI_{min}$ }         & {\bfseries $NI_{avg}$ }     & {\bfseries $ICA_{min}$  } & {\bfseries  $ICA_{avg}$ }  & {\bfseries $IO_{min}$ }     & {\bfseries $IO_{avg}$  } \\
			\hline
			 LRDS& RP & 0.8750&1.4490&0.4057&0.6942&0.0945&0.2932\\
			 & GP & 1.3248&1.6175&0.6402&0.7122&0.0584&0.4314\\
			 & PABIDOT &1.4046 & 1.4146&0.7038 & 0.7074& 0.6982&0.7067\\
			PBDS& RP & 0.7261&1.3368&0.5560&0.6769&0.0001&0.1242\\
			 & GP &0.2845&1.4885&0.1525&0.6834&0.0000&0.1048\\
			& PABIDOT & 1.4102 & 1.4166 & 0.6951 & 0.7042 & 0.6755 & 0.7032\\  
			SSDS& RP & 1.2820&1.5015&0.1751&0.5909&0.0021&0.0242\\
			 & GP & 1.4490&1.6285&0.0062&0.3240&0.0011&0.0111\\
			& PABIDOT&1.4058&1.4129&0.7069&0.7078&0.7031&0.7064\\
			WCDS & RP &1.0105&1.3098&0.6315&0.7362&0.0000&0.0895 \\
			& GP & 1.4620&1.7489&0.1069&0.6052&0.0000&0.1003\\
			& PABIDOT &1.3680&1.4260&0.6771&0.7051&0.6512&0.6809\\
			WQDS& RP & 1.2014&1.4957&0.4880&0.7062&0.0057&0.4809 \\
			 & GP & 1.3463&1.6097&0.3630&0.6536&0.0039&0.4025
 \\
            & PABIDOT&1.4019&1.4162&0.7034&0.7070&0.6901&0.7071 \\
            
            FRDS & RP & 1.2512 & 1.4318 & 0.6642  & 0.7129  & 0.3801 & 0.5270   \\
			 & GP & 1.4407 & 1.5675 & 0.6223 & 0.6965 & 0.3799 & 0.5475 
 \\
            & PABIDOT& 1.4057 & 1.4155 & 0.7051 & 0.7067 & 0.7019 & 0.7055 \\
            
            ELDS & RP & 1.0471 & 1.4001 & 0.6701 & 0.7282 & 0.0772 & 0.5554 \\
			 & GP & 1.4255 & 1.6036 & 0.7125 & 0.7587 & 0.0747 & 0.4952 
 \\
            & PABIDOT& 1.4079 & 1.4135 & 0.7003 & 0.7068 & 0.6991 & 0.7048\\
			\midrule
			\multicolumn{2}{c}{\bfseries FMR Values}& \multicolumn{2}{c}{RP: 1.65}& \multicolumn{2}{c}{GP: 1.85}& \multicolumn{2}{c}{PABIDOT: 2.50}\\
            \hline
		\end{tabular}
        }
	
	\end{small} 
	

\end{table}

The mean ranks produced by Friedman's rank test are presented in the last row of Table \ref{attackresistance}, with the following test statistics: $\chi^2$ value of 16.6108, a degree of freedom of 2 and a p-value of 2.4718e-04. The p-value suggests that the difference between the attack resistance values is significant. The mean ranks suggest that PABIDOT provides comparatively higher security against the privacy attacks.

\subsubsection{Privacy Guarantee}
\begin{figure}[H] 
	
	\centering
	\subfloat[Variation of $\phi$ of different axes against $\theta$.]{\includegraphics[width=0.49\textwidth, trim=0cm 0cm 0cm 0cm]{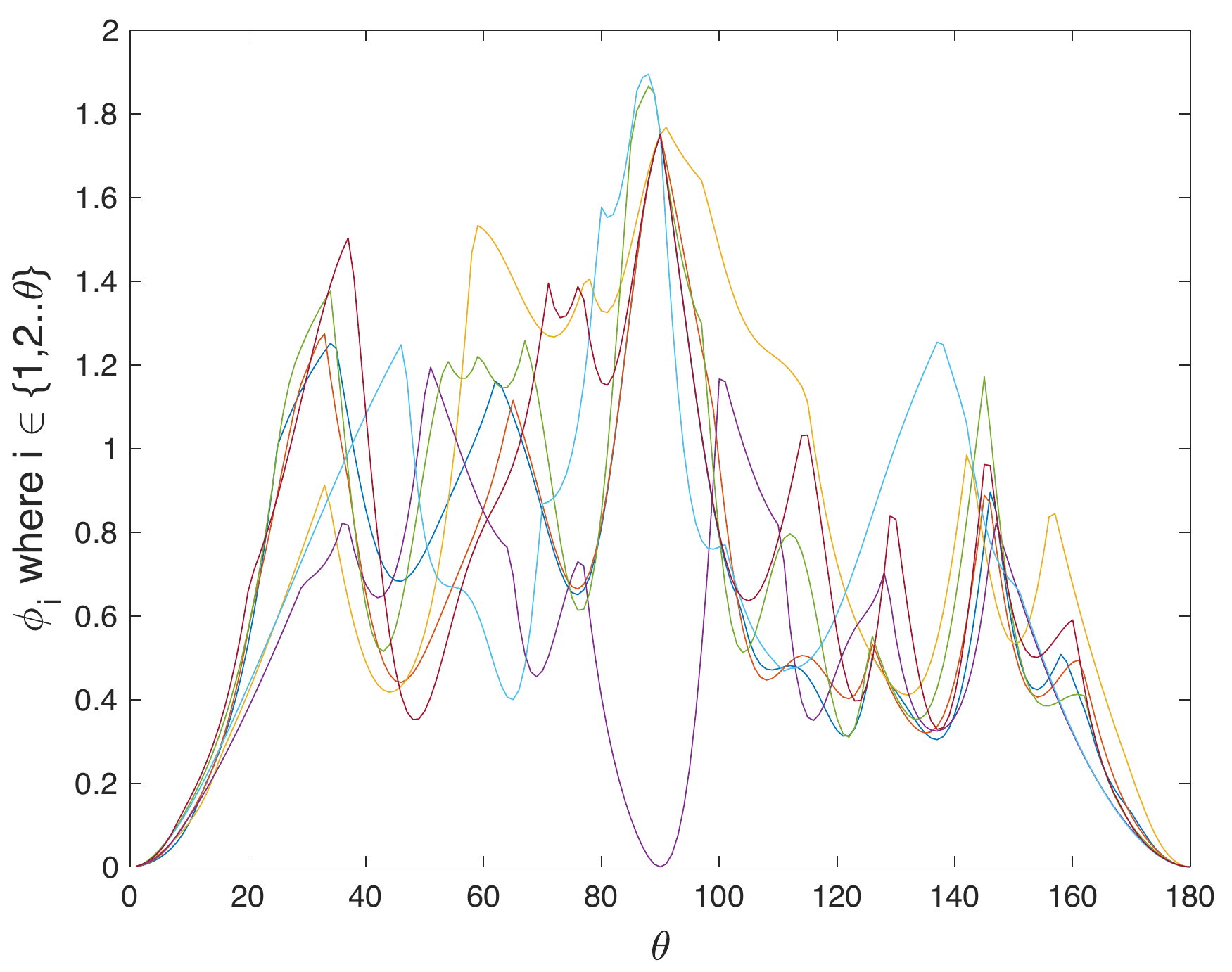}\label{theta_i_variation}}
	\hfill
	\subfloat[Variation of global $\phi$ against $\theta$.]{\includegraphics[width=0.47\textwidth, trim=0cm 0cm 0cm 0cm]{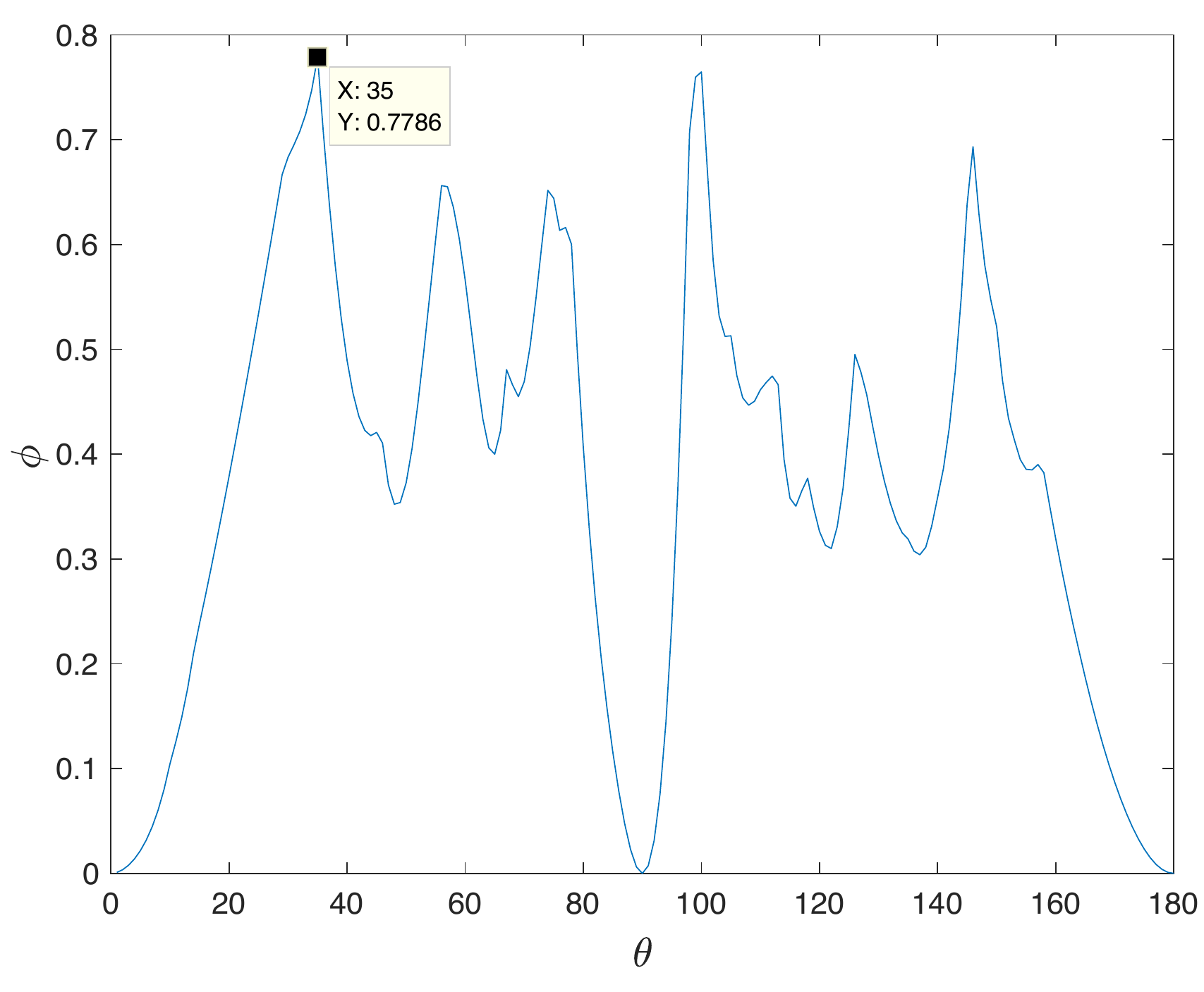}\label{theta_variation}}    
	\caption{$\phi$ vs. $\theta$. Figure \ref{theta_i_variation} shows variation of the local minimum privacy guarantee ($\phi_i$) curves for each attribute of the WCDS dataset. The $\phi_i$ values are utilized to generate the global minimum privacy guarantee ($\phi$) curve as shown in Figure \ref{theta_variation}; PABIDOT considers the global  maximum of $\phi$ to select the best perturbation parameters. For the WCDS dataset the best perturbation parameters are $\theta_{optimal}=35$ and $Rif_{optimal}=4$ which are taken at $\Phi=0.7786$.}
    \label{phi_vs_theta}
    

\end{figure}

\begin{figure}[H] 
	
	\centering
	\subfloat[minimum std(D-D$^r$) of the reconstructed datasets produced by the reconstruction attacks against $\theta$.]{\includegraphics[width=0.48\textwidth, trim=0cm 0cm 0cm 0cm]{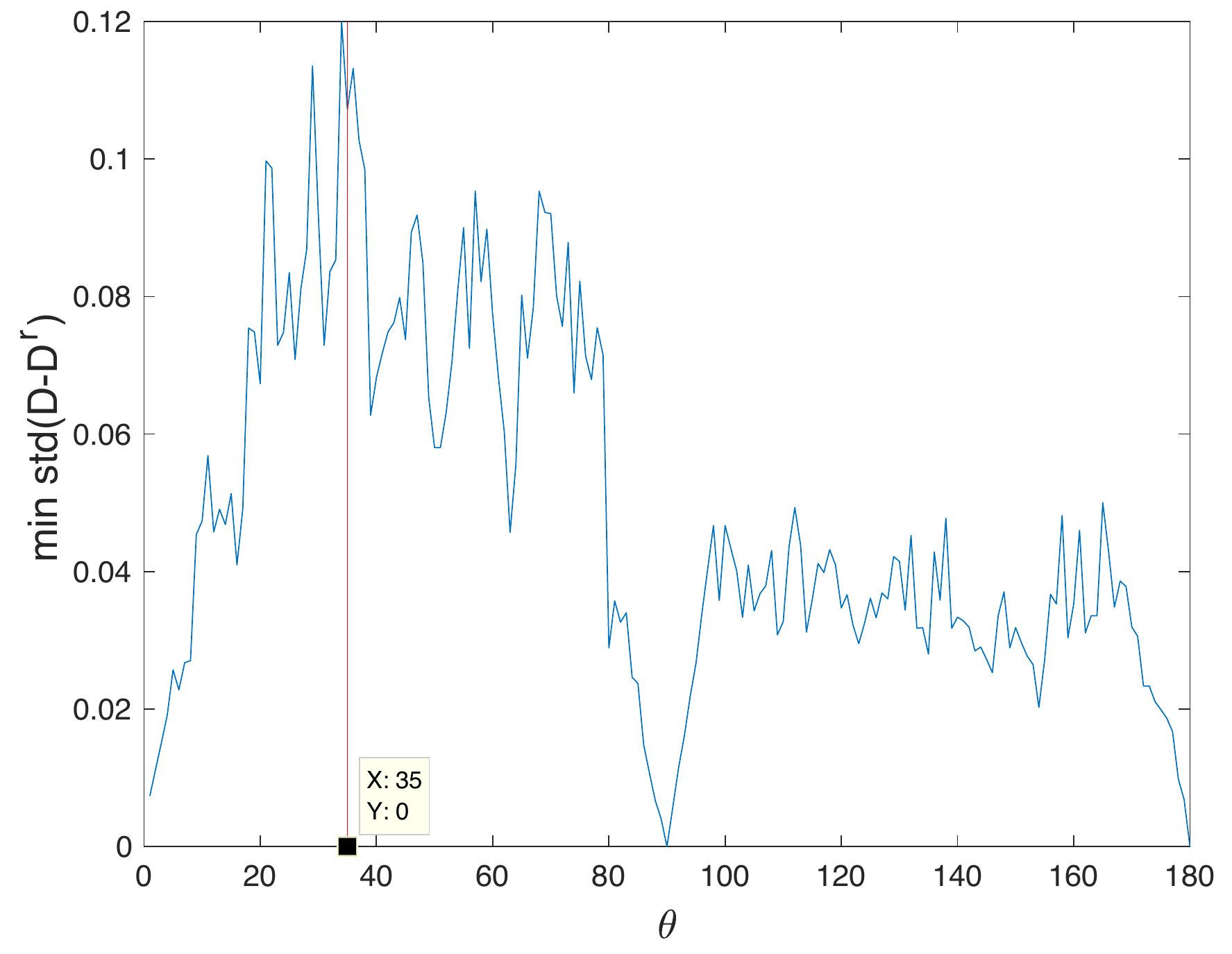}\label{phi_min_std}}
	\hfill
	\subfloat[average std(D-D$^r$) of the reconstructed datasets produced by the reconstruction attacks against $\theta$.]{\includegraphics[width=0.48\textwidth, trim=0cm 0cm 0cm 0cm]{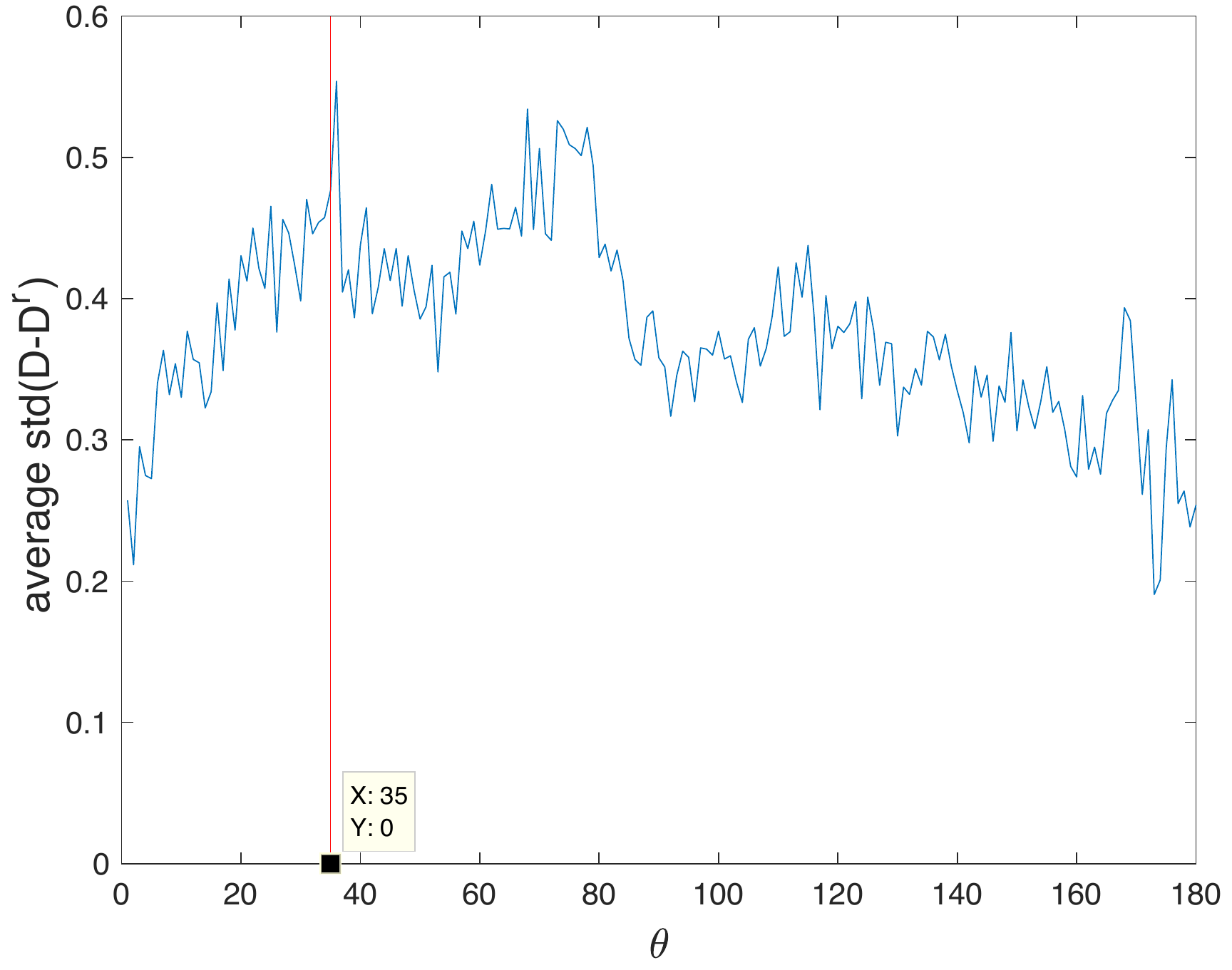}\label{phi_avg_std}}    
	\caption{minimum std(D-D$^r$) and average std(D-D$^r$) of the reconstructed datasets produced by Naive Snooping, ICA and known I/O. The red vertical lines show the instance of optimal perturbation parameter selection of PABIDOT. The red lines nearly indicate the point at which the corresponding perturbed dataset provides the highest privacy guarantee. This provides empirical evidence on PABIDOT providing the optimal privacy by satisfying $\Phi-separation$.}
    \label{phi_attack_std}
    

\end{figure}

To determine the privacy guarantee of the $\Phi-separation$ model, we investigated the dynamics of $\Phi$ against PABIDOT's resistance to data reconstruction attacks. In this experiment, we used the WCDS  dataset as it is the smallest and most vulnerable dataset against data reconstruction attacks (refer to Table\ref{attackresistance}). Figure \ref{theta_i_variation} shows the resulting $\phi_i$ (local minimum privacy guarantee) curves for the different attributes (considered as axes in PABIDOT) of WCDS. The minimum of the $\phi$ values available in each $\phi_i$ curve under each $\theta$ are obtained to generate the plot of global minimum privacy guarantees as shown in Figure \ref{theta_variation}. Next, the global maximum of $\phi$ values is selected as $\Phi$. As shown on Figure \ref{theta_variation}, the best perturbation parameters for the WCDS dataset are, $\theta_{optimal}=35$ and $Rif_{optimal}=4$ which are taken at $\Phi=0.7786$.  
In order to determine whether the selected perturbation parameters, that were determined based on $\Phi-separation$, actually provide enough privacy guarantee, we generated perturbed datasets for each instance of $\theta$ with the best $axis$  $of$ $reflection$ under the corresponding $\theta$ values. To determine the absolute impact of $\Phi$ on the analysis, the effect of randomized expansion and random tuple shuffling  were minimized by selecting an input $\sigma$ value of 0 and finally sorting the resulting datasets in ascending order. Next, naive snooping, ICA and known I/O attack methods were used to reconstruct the datasets. Then, the reconstructed datasets were used to produce  the $min$ $std(D-D^r)$ and $average$ $std(D-D^r)$ under each $\theta$ instance, where $D^r$ represents the reconstructed datasets.  Next, $min$ $std(D-D^r)$ and $average$ $std(D-D^r)$  were plotted against $\theta$ as shown in Figure \ref{phi_min_std} and Figure \ref{phi_avg_std}  respectively. The red colored vertical line indicates the point of $\Phi$. As shown in Figures \ref{phi_min_std} and \ref{phi_avg_std}, it's clear that $\Phi$ provides a near optimal point at which PABIDOT produces the highest  $min$ $std(D-D^r)$ and $average$ $std(D-D^r)$.  This in fact proves that $\Phi-separation$ provides an empirical privacy guarantee against data reconstruction attacks.

\begin{figure}[H] 
	
\centering
\subfloat[Minimum $std(D-D^r)$ vs. standard deviation of noise.]{\includegraphics[width=0.48\textwidth, trim=0cm 0cm 0cm 0cm]{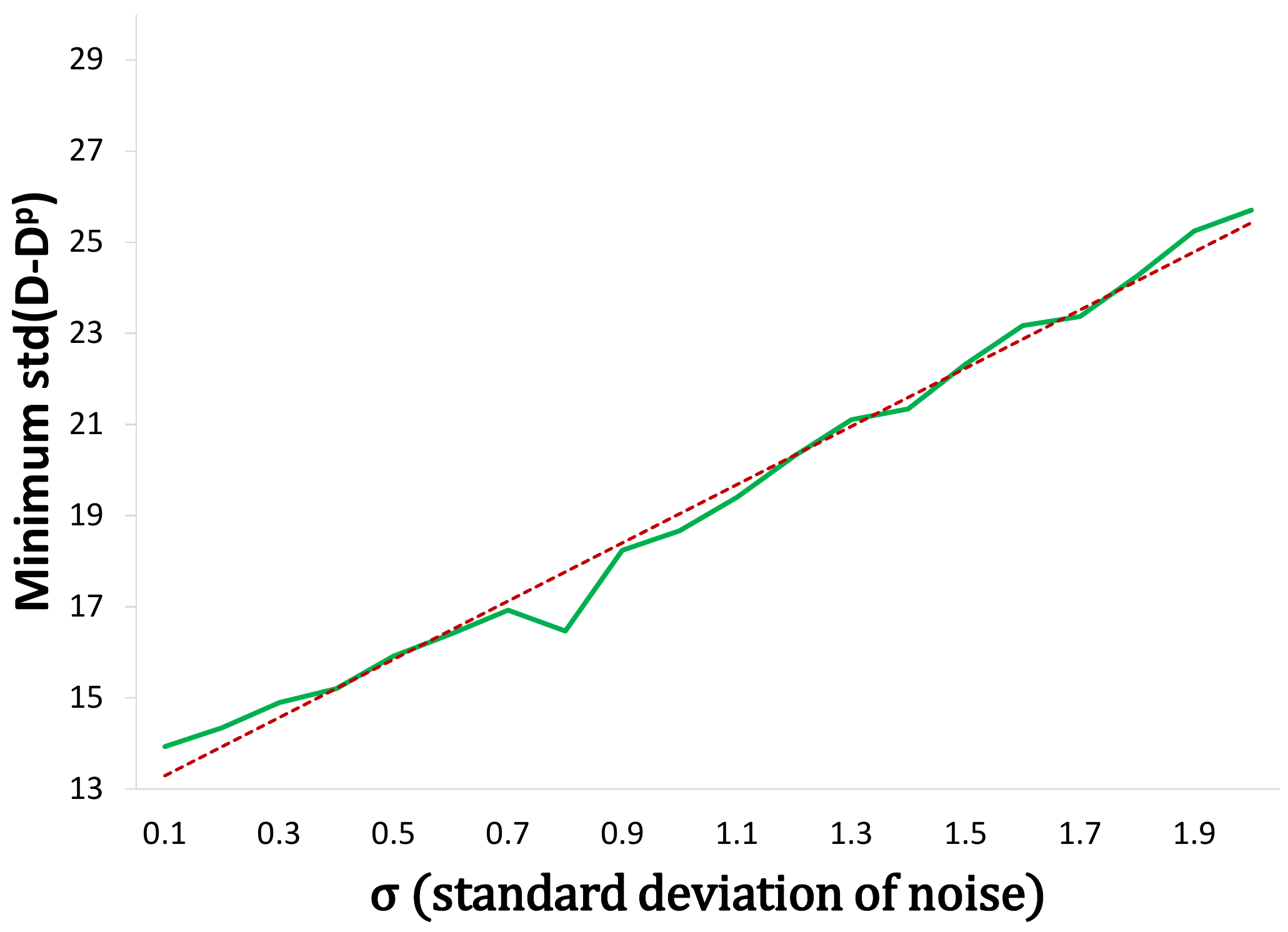}\label{composition_stddif}}
	\hfill
	\subfloat[Classification accuracy vs. standard deviation of noise.]{\includegraphics[width=0.48\textwidth, trim=0cm 0cm 0cm 0cm]{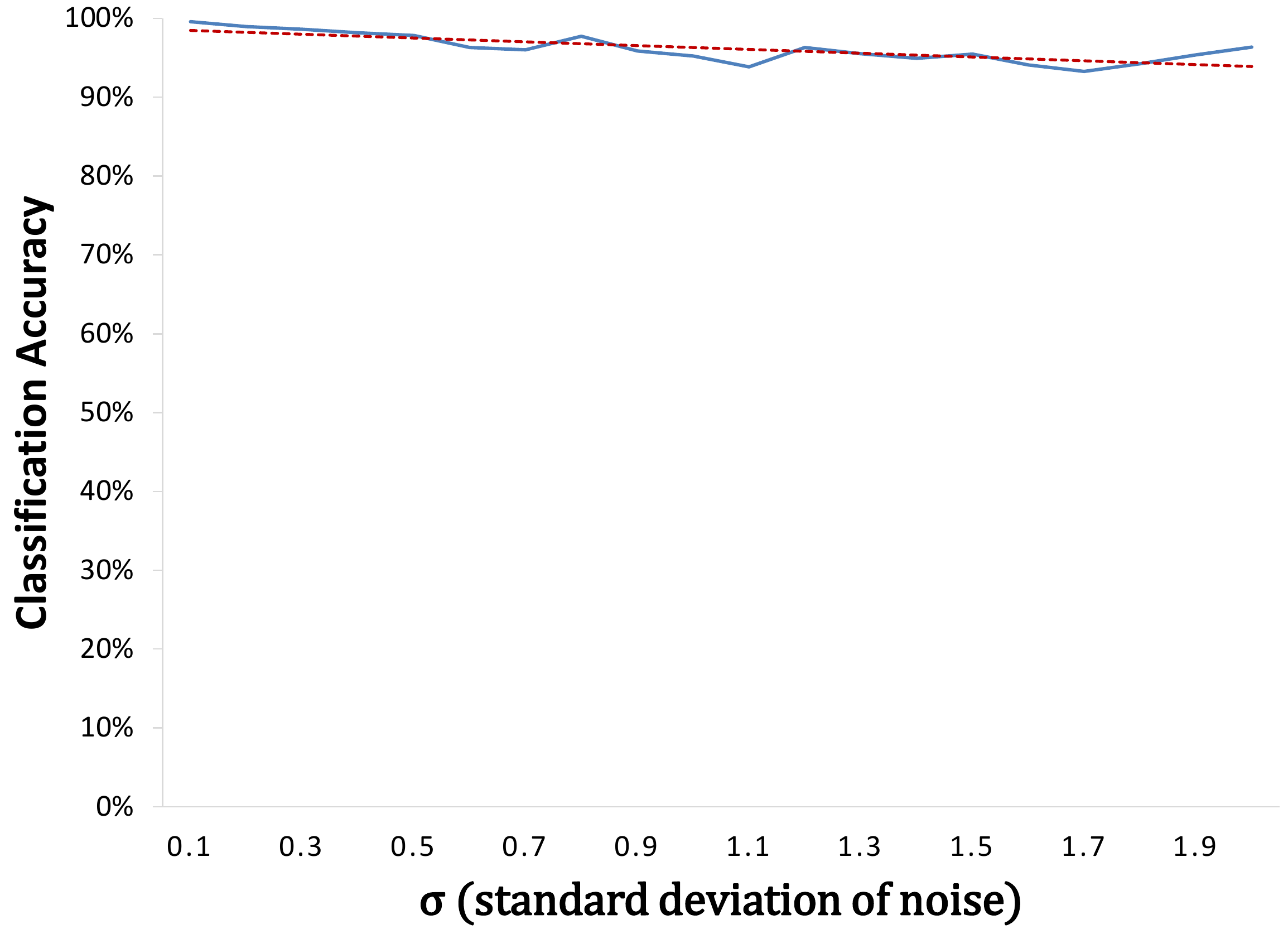}\label{composition_accuracy}}    
\caption{Effect of $\sigma$ on $min(std(D-D^r))$ and classification accuracy. When the $\sigma$ of the randomized expansion is increased, the $minimum$ $std(D-D^p)$ increases as shown in Figure \ref{composition_stddif}. However, the classification accuracy shows only a minimal decrease against increasing $\sigma$. This confirms PABIDOT's capability of maintaining utility at a constant level while providing increased resistance to increasing randomized expansion.}
 \label{composition}
 

\end{figure}

\paragraph*{\textbf{Composition of PABIDOT}} PABIDOT performs steps \ref{al2step22} and \ref{al2step24} (refer Algorithm \ref{parallelalgo}) to provide composition and resistance to complementary release attacks which exploit the vulnerabilities of multiple perturbed data releases.  Figure \ref{composition_stddif} shows the change of $std(D-D^p)$ against the standard deviation ($\sigma$) of noise used at step \ref{al2step22}. The results were produced using the SSDS dataset and IBK classification algorithm. In a complementary data release, the data owner is opted to release several perturbed versions of the dataset consecutively. In each release, the data owner will have to increase $\sigma$ in order to increase the deviation of the data by increasing the noise of the dataset. The random tuple swapping (conducted in step \ref{al2step24}) makes the process of direct data linkage unrealistic. As shown in Figure \ref{composition_accuracy}, the classification accuracy does not get reduced drastically with increasing $\sigma$.

\paragraph*{\textbf{Membership inference attacks}} Membership inference attacks (MI) investigate whether a record was a member of the training dataset of a model with black-box access. This attack is used to investigate information leaks in models trained by commercial ``machine learning as a service" providers such as Google and Amazon.~\cite{shokri2017membership}. In our proposed work, we work on full data release as depicted in Figure \ref{algoflow}. We assume that only the perturbed data is released and the original data is not accessible by any third party user. Therefore, PABIDOT is not directly aligned with the concept of any model release with black box access to them. Nevertheless, it can be important to infer the membership of the data perturbed by PABIDOT, in a model trained with original data. This investigation is saved for our future work.

\subsubsection{Increase in Information Entropy}
The information entropy of the perturbed and the non-perturbed datasets were compared by using Shannon's entropy formula given in Equation \ref{shannon} \cite{park2003anomaly}. 

\begin{equation}
H(P)=-\sum_{i=1}^{k}p_ilog_2p_i
\label{shannon}
\end{equation}

\begin{table}[H]
	\centering
	
	\caption{Average Increase in Information Entropy. The average increase in entropy of perturbed data compared to the original data was calculated using Equation \ref{averageshannon}. The positive values show that there is always an increase in entropy, and that shows that PABIDOT introduces more impurity to the perturbed data.} 
	\label{information}
	\setlength\tabcolsep{5pt} 
	\begin{small}
        \resizebox{0.6\columnwidth}{!}{
		\begin{tabular}{ l l }
			\hline
			{\bfseries Dataset} & {\bfseries  Average increase in information entropy} \\		
			\hline
			LRDS & 11.1933 \\
			PBDS & 4.1284 \\ 
			SSDS & 11.2074 \\ 
			WCDS & 1.1671\\
			WQDS & 5.9415 \\ 
			FRDS & 5.3300 \\ 
			ELDS &  7.2458\\ 
			\hline
		\end{tabular}
        }
	\end{small} 
	

\end{table}

Table \ref{information} shows the average of the increase in information entropy obtained according to Equation \ref{averageshannon}. The average increase is positive in all cases, proving that the attributes of the perturbed datasets have higher information entropies, hence having more impurity compared to original data.

\begin{equation}
AIG(D^p-D)=\frac{\sum_{i=1}^{k}(H(P_i^p)-H(P_i))}{k}
\label{averageshannon}
\end{equation}

\section{Discussion}
\label{discussion}
In this paper, we proposed an efficient perturbation algorithm (named PABIDOT) for preserving individuals' privacy in the context of big data. PABIDOT applies perturbation based on a new privacy model name $\Phi-separation$. We proved that $\Phi-separation$ provides an empirical privacy guarantee of selecting the best perturbation parameters.  PABIDOT excels in time efficiency, and the classification accuracy results of the perturbed dataset are close to that of the original data. This was achieved by taking a systematic approach to optimizing data perturbation parameters for random axis reflection, noise translation, and multidimensional concatenated sub-plane rotation followed by the introduction of novel random noise addition mechanism named as randomized expansion. The utility of  PABIDOT was analyzed in terms of time complexity, scalability, memory consumption, classification accuracy and biases with post-processing properties. Next, the privacy of PABIDOT was analyzed in terms of attack resistance, privacy guarantee, composition, and increase in information entropy.

In data classification experiments using three data perturbation methods, PABIDOT outperformed two related methods: rotation perturbation (RP) and geometric perturbation (GP). Friedman's rank test further supports this argument by returning the highest mean rank for PABIDOT's classification accuracy results. The classification accuracy of the dataset perturbed by PABIDOT was second only to that of the original dataset.  

The time complexity of PABIDOT was investigated through analysis and experimentation and is given as $O(n^3\times m)$ where $n$ is the number of attributes, and $m$ is the number of tuples in the dataset. Conventionally a dataset grows by the addition of new tuples, while the number of attributes remains constant. In case of large datasets, we can consider an inequality between $m$ and $n$ as $(m>>>n)$, and so the time complexity of PABIDOT is linear ($O(m)$) for $m$, i.e. when the number of attributes remains constant. This indicates the suitability of the proposed perturbation method for big data. The time consumption experiments exhibit the same pattern, as depicted in Figure \ref{timecompthree}, and agree with the time complexity analysis.

Scalability testing also gave good results; runtime analysis showed that PABIDOT completes the perturbation process much faster than RP or GP. In fact, RP and GP did not converge within the maximum time (100 hours) given for an individual job in the SGI UV3000 supercomputer. This proves that the new method is usable for large-scale data perturbation with very low computational overheads.

Memory requirements of PABIDOT are moderate, in our experiments PABIDOT and rotation perturbation (RP) used less memory than geometric perturbation (GP). According to Friedman's rank test, PABIDOT uses less memory than RP, but the p-value suggests that the difference is not significant. We can conclude that the proposed method works well with datasets that have a substantial number of tuples and a large number of attributes.

It was concluded that perturbation alters the descriptive statistics, making the resulting data to be unusable for descriptive statistical analysis. This also suggests that an adversary cannot try to conduct database inference attacks based on descriptive statistics. PABIDOT has lower levels of type B and type C biases, which means that PABIDOT does not affect the relationship between the attributes. Further analysis of biases proved that the probability distributions between the original and perturbed attributes are affected, meaning that direct mapping between the attributes is not possible.

The privacy guarantee of $\Phi-separation$ makes sure that PABIDOT perturbs the dataset with optimal perturbation parameters for a given instance. The proposed method of randomized expansion further reduces the possibility of reconstructing the original data. In attack resistance experiments, PABIDOT showed better protection against attacks that try to restore the original dataset than  RP and GP. PABIDOT applies reverse z-score normalization after the transformations, so the final attributes' value ranges are within those of the original dataset. This can reduce the probability of attacks in the first place, as attackers may not be able to distinguish the original dataset from the perturbed one.

PABIDOT resists various attacks as shown in Table \ref{risklevel}. In particular, it shows a low level of vulnerability to ICA based, Known I/O, and Naive inference attacks. Since PABIDOT performs randomized expansion and random shuffling of the resulting dataset, multiple releases would have only a minor effect on the privacy it provides. 
Hence, complementary release attacks that exploit the privacy loss due to multiple releases do not represent a serious risk. This provides an empirical guarantee of the composability of PABIDOT with the assumption that the data owner increases the noise factor ($\sigma$) in each sequential data release. Since PABIDOT does not try to minimize information loss over generalization, minimality attacks ~\cite{wong2007minimality} do not pose a risk either. In the experiments section, we pointed out that membership inference has a considerably low level of applicability to PABIDOT. So we can claim that PABIDOT  is practically not vulnerable to membership inference attacks.  Distance inference attacks and Eigen-Analysis are still being investigated, hence, indicated as inconclusive.

\begin{table}[H]
  \centering
  \caption{resistance of the proposed method against existing attacks}
  \resizebox{0.6\columnwidth}{!}{

    \begin{tabular}{l c c c c}
    \toprule
    \multirow{2}[4]{*}{\textbf{Attack \newline{}Method}} & \multicolumn{4}{c}{\textbf{Risk Level}} \\
\cmidrule{2-5}    \multicolumn{1}{c}{} & \textbf{Inconclusive} & \textbf{Low} & \textbf{Moderate} & \textbf{High} \\
    \midrule
    Complementary Release &       &   \checkmark   &       &   \\
    ICA based &       & \checkmark     &       &  \\
    Known I/O &       & \checkmark     &       &  \\
    Naive Estimation &       & \checkmark     &       &  \\
    Minimality &       &       &       &  \\
    Membership Inference &       & \checkmark     &       &  \\
    Distance-inference & \checkmark     &       &       &  \\
    Known Sample &      &    \checkmark    &      &  \\
    Eigen-Analysis & \checkmark     &       &       &  \\
    \bottomrule
    \end{tabular}%
    }
  \label{risklevel}%
\end{table}

Finally, the increase in entropy of the perturbed data was investigated. Higher entropy indicates increased impurity in a dataset,  that is, the higher the entropy, the higher the number of different records ~\cite{park2003anomaly}. Consequently, increasing the entropy of a dataset makes the process of reconstructing the original data more difficult. PABIDOT performs well in this respect; it increases the entropy as given in Table \ref{information}.

In summary, PABIDOT's suitability for the privacy preservation of big data was demonstrated from several perspectives. Data perturbation is a low-cost way of preserving privacy, and among data perturbation methods, PABIDOT proved to be very fast. PABIDOT scales well with linear complexity for typical scenarios when the number of attributes is a constant.  PABIDOT's memory usage is comparable to that of the existing  methods. PABIDOT also showed better resistance to attacks on privacy than comparable techniques. There are many practical examples in different domains which produce big data that contain various private information. For example, healthcare systems maintain large amounts of patient data to improve clinical analytics and deliver efficient services to medical staff and patients ~\cite{manogaran2017big}. In retail environments, an enormous amount of data are created through customer transactions~\cite{aloysius2018big}. These retail data are vital to seek initiatives to leverage customer services. However, data stored in these systems include private information that needs protection. If the data are not appropriately looked after, unanticipated privacy leaks may occur. One of the significant issues in enforcing privacy is the high dimensions of these data. PABIDOT provides a scalable and efficient solution that can be effectively used under such challenging environments to preserve the confidentiality of user data. 

Another example is Facebook that has a massive app base, and exploitation of vulnerabilities can have devastating consequences wrt users' privacy~\cite{steel2010facebook}.  Using a robust privacy preserving mechanism such as PABIDOT could efficiently address these issues. In fact, events like Facebook's privacy breaches could have been avoided using a perturbation mechanism such as PABIDOT~\cite{steel2010facebook}. Hence, PABIDOT would not only preserve individual user privacy but also allow other third-party companies to generate population-based valuable insights without harming the privacy of users. 

\section{Conclusion}
\label{conclusion}
This paper proposed a new perturbation method named PABIDOT in order to address efficiency, scalability, privacy, usability and utility issues of the existing data perturbation methods. PABIDOT adds privacy to data by using a newly proposed privacy model named $\Phi-separation$. We provided empirical evidence that $\Phi-separation$ provides sufficient privacy guarantee by efficiently selecting the best perturbation parameters for PABIDOT's data perturbation. The asymptotic complexity of PABIDOT was proven to be linear ($O(m)$) for $m$ where $m$ is the number of instances and $m$ is considerably larger than the number of attributes for a given dataset (the number of attributes is often a constant for a particular dataset). This property enables PABIDOT to deal with large datasets. The scalability results proved that PABIDOT consumes an extremely low amount of time for the perturbation of big data. The classification accuracy of the proposed method is very close to that of the original dataset, and the method outperforms random rotation perturbation and geometric perturbation in this respect. The composition and privacy guarantee of the proposed algorithm are additional advantages. Empirical results show that PABIDOT provides good resistance to various privacy attacks. 

Further enhancing the efficiency and scalability of PABIDOT for big datasets can be a new research avenue.

\end{document}